\documentclass[preprintnumbers,amsmath,amssymb]{revtex4}
\usepackage{dcolumn}
\usepackage{bm}
\usepackage[dvips]{color}
\usepackage[pdftex]{graphicx}

\newcommand{\beq}{\begin{equation}}
\newcommand{\eeq}{\end{equation}}
\newcommand{\bq}{\begin{equation}}
\newcommand{\eq}{\end{equation}}
\newcommand{\be}{\begin{equation}}
\newcommand{\ee}{\end{equation}}
\newcommand{\ba}{\begin{array}}
\newcommand{\ea}{\end{array}}
\newcommand{\beqa}{\begin{eqnarray}}
\newcommand{\eeqa}{\end{eqnarray}}
\newcommand{\bea}{\begin{eqnarray}}
\newcommand{\eea}{\end{eqnarray}}

\def\ol{\overline}

\def\[{\left[}
\def\]{\right]}
\def\({\left(}
\def\){\right)}
\def\dif{\partial}

\evensidemargin=\oddsidemargin

\def\U{\mathcal{U}}
\def\OBZ{\mathcal{O}_{\mathcal{BZ}}}
\def\OU{\mathcal{O}_\mathcal{U}}
\def\BZ{\mathcal{BZ}}
\def\OSM{\mathcal{O}_{\rm SM}}
\def\g5{\gamma_5}

\def\pslash{\not{\hbox{\kern-4pt $p$}}}
\def\qslash{\not{\hbox{\kern-4pt $q$}}}
\def\lv{\not{\hbox{\kern-4pt $L$}}}
\def\lsim{\mathrel{\raise.3ex\hbox{$<$\kern-.75em\lower1ex\hbox{$\sim$}}}}
\def\gsim{\mathrel{\raise.3ex\hbox{$>$\kern-.75em\lower1ex\hbox{$\sim$}}}}
\def\ifmath#1{\relax\ifmmode #1\else $#1$\fi}
\begin{document}

 \title{\large The large CP phase in  
$B_{s}-\ol{B}_{s}$ mixing and Unparticle Physics}

 \author{J.\ K.\ Parry}~

 \affiliation{
       Center for High Energy Physics,
       Tsinghua University, Beijing 100084, China\,}
 \email{jkparry@tsinghua.edu.cn}

 \begin{abstract}
In this work we investigate the contribution to $B_{d,s}$ mixing 
from both scalar and vector unparticles in a number of scenarios. 
The emphasis of this work is
to show the impact of the recently discovered $3\sigma$ evidence for 
new physics found in the CP phase of $B_s$ mixing. Here we show that the 
inclusion of the CP phase constraints for both $B_d$ and $B_s$
mixing improves the bounds set on the unparticle couplings by a factor
of $2\sim 4$, and one particular scenario of scalar unparticles 
is found to be excluded by the $3\sigma$ measurement of $\phi_s$. 
 \\[2mm]
  PACS: 13.20.He  13.25.Hw  12.60.-i    \hfill
  [\,{\tt arXiv:0806.4350}\,]
 \end{abstract}
 \maketitle

\section{Introduction}

It has recently been suggested \cite{hep-ph/0703260,arXiv:0704.2457} that there
may exist a nontrivial scale invariant sector at high energies, 
known as unparticle stuff. These new fields with an 
infrared fixed point are called Banks-Zaks fields \cite{Nucl.Phys.B196.189},
interacting with standard model fields via heavy particle exchange,
\bea
\frac{1}{M_\U^k}\OSM\OBZ.
\eea
Here $\OSM$ is a standard model (SM) operator of mass dimension $d_{\rm SM}$,
$\OBZ$ is a Banks-Zaks($\mathcal{BZ}$) operator of mass dimension 
$d_{\mathcal{BZ}}$, with $k=d_{\rm SM}+d_{\mathcal{BZ}}-4$,
 and $M_\U$ is the mass of the heavy particles mediating
the interaction. At a scale denoted by $\Lambda_\U$ the $\BZ$ operators
match onto unparticle operators with a new set of interactions,
\bea
C_\U \frac{\Lambda_\U^{d_\BZ-d_\U}}{M_\U^k}\OSM\OU
\eea
where $\OU$ is an unparticle operator with scaling dimension $d_\U$ 
and $C_\U$ is the coefficient of the low-energy theory. Unparticle stuff
of scaling dimension $d_\U$ looks like a nonintegral number $d_\U$ of
invisible massless particles. In this work we shall study both 
scalar unparticles ($\OU$) and vector unparticles ($\OU^\mu$ ) 
with couplings to the SM quarks as follows,
\bea
{\rm Scalar\, unparticles:}&\hspace{1cm}&
\frac{c_{L}^{S,\,q'q}}{\Lambda_\U^{d_\U}}\,
\ol{q}'\gamma_\mu(1-\g5)q\,\dif^\mu\,\OU
+\frac{c_{R}^{S,\,q'q}}{\Lambda_\U^{d_\U}}\,
\ol{q}'\gamma_\mu(1+\g5)q\,\dif^\mu\,\OU\label{S_int}\\
{\rm Vector\, unparticles:}&\hspace{1cm}&
\frac{c_{L}^{V,\,q'q}}{\Lambda_\U^{d_\U-1}}\,
\ol{q}'\gamma_\mu(1-\g5)q\,\OU^\mu
+\frac{c_{R}^{V,\,q'q}}{\Lambda_\U^{d_\U-1}}\,
\ol{q}'\gamma_\mu(1+\g5)q\,\OU^\mu\
\label{V_int}
\eea 
Here we assume that the left-handed and right-handed 
flavour-dependent dimensionless
couplings $c^S_{L}$, $c^S_{R}$, $c^V_{L}$ and $c^V_{R}$ 
are independent parameters.
We analyze a number of scenarios, in each case determining the 
allowed parameter space and placing bounds on the unparticle couplings.

The propagators for scalar and vector unparticle fields are as
follows \cite{hep-ph/0703260,arXiv:0704.2588},
\bea
\int\,d^4 x \, e^{iP.x}
\langle 0| T\OU(x)\OU(0)|0\rangle&=&
i\frac{A_{d_\U}}{2\sin d_\U\pi}
\frac{1}{(P^2+i\epsilon)^{2-d_U}}e^{-i\phi_\U}\\
\int\,d^4 x \, e^{iP.x}
\langle 0| T\OU^\mu(x)\OU^\nu(0)|0\rangle&=&
i\frac{A_{d_\U}}{2\sin d_\U\pi}
\frac{(-g^{\mu\nu}+P^\mu P^\nu/P^2)}{(P^2+i\epsilon)^{2-d_U}}e^{-i\phi_\U}
\eea
where
\bea
A_{d_\U}=\frac{16\pi^{5/2}}{(2\pi)^{2d_\U}}
\frac{\Gamma(d_\U+1/2)}{\Gamma(d_\U-1)\Gamma(2d_\U)},
\hspace{0.5cm}
\phi_\U=(d_\U-2)\pi
\eea
Since the publication of the first theoretical papers on this new
subject \cite{hep-ph/0703260,arXiv:0704.2457} there has been a huge interest
in unparticle phenomenology, for example, Collider signatures 
\cite{arXiv:0704.2588,arXiv:0704.3532,arXiv:0705.0794,arXiv:0705.2622,arXiv:0705.3518,arXiv:0705.4542,arXiv:0705.4599,arXiv:0706.1284,arXiv:0706.3155,arXiv:0707.0187,arXiv:0707.0893,arXiv:0707.2074,arXiv:0708.0036,arXiv:0708.0671,arXiv:0708.3802,arXiv:0708.1960,arXiv:0709.2478,arXiv:0709.1505,arXiv:0710.2230,arXiv:0710.4239,arXiv:0710.5773,arXiv:0711.1665,arXiv:0711.3361,arXiv:0805.1150,arXiv:0805.0199,Mureika:2007nc,Alan:2007rg,liaoy1,liaoy2,liaoy3,liaoy4,liaoy5},
CP violation
\cite{arXiv:0705.0689,arXiv:0706.0850,arXiv:0707.1268,arXiv:0709.0235,Zwicky:2007yc,Zwicky:2007vv},
meson mixing 
\cite{arXiv:0705.1821,arXiv:0704.3532,arXiv:0705.0689,arXiv:0707.1234,arXiv:0707.1535,arXiv:0710.3663,arXiv:0711.3516,arXiv:0801.0895},
lepton flavour violation
\cite{arXiv:0705.1326,arXiv:0705.2909,arXiv:0709.3435,arXiv:0711.2744,arXiv:0806.2944,arXiv:0802.4015,arXiv:0802.1277},
consequences in astrophysics
\cite{arXiv:0705.3636,arXiv:0708.1404,arXiv:0708.2812,arXiv:0708.4339,arXiv:0710.4275,arXiv:0711.1506,arXiv:0803.3223,liaoy6},
in neutrino physics
\cite{arXiv:0706.0302,arXiv:0706.0325,arXiv:0707.2285,arXiv:0708.3485,arXiv:0709.0678}
and in supersymmetry
\cite{arXiv:0707.2132,arXiv:0707.2451,arXiv:0707.2959}.
In this work we shall study the constraints coming from the measurements
of $B_{s,d}$ meson mass differences $\Delta M_{s,d}$ and also their 
CP violating phases $\phi_{s,d}$. In the $B_d$ system these quantities have
been well measured for some time and show only small deviations from the 
SM expectation. In the $B_s$ system recent measurements have also found
small discrepancies between the SM expectation for $\Delta M_s$ 
\cite{hep-ex/0609040}, but now
the CP violating phase $\phi_s$, measured by the 
${\rm D}\emptyset$\cite{arXiv:0802.2255} 
and CDF\cite{arXiv:0712.2397} collaborations
reveals a deviation of 3$\sigma$\cite{arXiv:0803.0659,Lenz:2006hd}
\footnote{It should be noted that the combination of D$\emptyset$ and 
CDF results for $\phi_s$ were made 
without full knowledge of the likelihoods, which 
could have an effect on the significance of this SM deviation.
These are now being made available and a new experimental combination
is due to be released soon. Until that time this present hint of new physics 
is the best available and remains a very exciting prospect. }. 
This is the first evidence for new physics
in $b\leftrightarrow s$ transitions. Studying both vector 
and scalar unparticles, we 
study the constraints imposed by these latest measurements on the 
coupling between SM fields and unparticles, with particular interest
on the impact of $\phi_s$.

In Sec.~\ref{sec2} we discuss unparticle contributions to
general meson-antimeson mixing and in particular, to the case of 
$B$ mixing. Section~\ref{sec3} contains the results of our 
numerical analysis where we first consider scalar and then vector
unparticle effects in $B_{d,s}$ mixing. In each case we set bounds
on the allowed parameter space using constraints from measurements
of $\Delta M_{d,s}$ and from $\phi_{d,s}$. Finally, Sec.~\ref{conc}
concludes our results.

\section{Meson-antimeson mixing from unparticles}\label{sec2}

\begin{figure}[h]
\begin{center}
\includegraphics[width=11truecm,clip=true,angle=0]{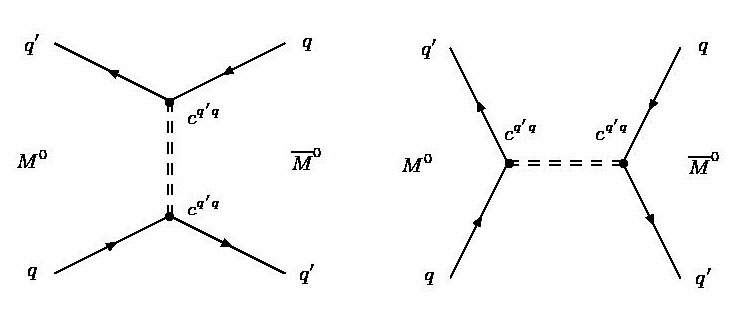}
\end{center}
\caption{$s$- and $t$-channel unparticle contributions to meson mixing.
\label{fig_mixing}}
\end{figure}

With unparticle operators coupling to standard sodel operators 
as in Eq.~(\ref{S_int},\ref{V_int}) it is possible for unparticle
physics to contribute to meson-antimeson mixing via the $s$- and $t$-channel
processes shown in Fig.~\ref{fig_mixing}.

Using the interactions listed in Eq.~(\ref{S_int},\ref{V_int})
to evaluate the s- and t-channel contributions to meson mixing as shown
in Fig.~\ref{fig_mixing} we obtain the effective Hamiltonian for 
scalar unparticles as,
\bea
\mathcal{H}^{S,q'q}_{\rm eff}&=&
\frac{A_{d_\U}}{2\sin d_\U\pi}\frac{e^{-i\phi_{\U}}}{\Lambda_\U^{2d_\U}}
\left(\frac{1}{t^{2-d_\U}}+\frac{1}{s^{2-d_\U}}\right)\nonumber\\
&\times& \left[
Q_{2}
\left(
m_{q'}c_L^{S,q'q}-m_{q}c_R^{S,q'q}
\right)
\left(
m_{q}c_R^{S,q'q}-m_{q'}c_L^{S,q'q}
\right)
\right.\nonumber\\
&&+\left.
\tilde{Q}_{2}
\left(
m_{q'}c_R^{S,q'q}-m_{q}c_L^{S,q'q}
\left)
\right(
m_{q}c_L^{S,q'q}-m_{q'}c_R^{S,q'q}
\right)
\right.\label{Hscal}\\
&&+\left.
2\,Q_{4}
\left(
m_{q'}c_L^{S,q'q}-m_{q}c_R^{S,q'q}
\right)
\left(
m_{q}c_L^{S,q'q}-m_{q'}c_R^{S,q'q}
\right)
\right]\nonumber
\eea
and for vector unparticles we obtain,
\bea
\mathcal{H}^{V,q'q}_{\rm eff}
&=&
\frac{A_{d_\U}}{2\sin d_\U\pi}\frac{e^{-i\phi_\U}}{\Lambda_\U^{2d_\U-2}}
\nonumber\\
\times&&\hspace{-0.6cm}
\left\{
\left(\frac{1}{t^{3-d_\U}}+\frac{1}{s^{3-d_\U}}\right)
\right.\left.
\left[
Q_2 
\left(
m_{q'}\,c_L^{V,q'q}-m_q \,c_R^{V,q'q}
\right)
\left(
m_q \,c_R^{V,q'q}-m_{q'}\,c_L^{V,q'q}
\right)
\right.\right.\nonumber\\
&&\hspace{2.8cm}+\left.\left.
\tilde{Q}_2
\left(
m_{q'}\,c_R^{V,q'q}-m_q \,c_L^{V,q'q}
\right)
\left(
m_q \,c_L^{V,q'q}-m_{q'}\,c_R^{V,q'q}
\right)
\right.\right.\nonumber\\
&&\hspace{2.8cm}+\left.\left.
2\,Q_4
\left(
m_{q'} \,c_L^{V,q'q}-m_q \,c_R^{V,q'q}
\right)
\left(
m_{q} \,c_L^{V,q'q}-m_{q'} \,c_R^{V,q'q}
\right)
\right]\right.\label{Hvec}\\
&-&\left.
\left(\frac{1}{t^{2-d_\U}}+\frac{1}{s^{2-d_\U}}\right)
\left[
Q_{1}\,\left(c_{L}^{V,q'q}\right)^2 
+\tilde{Q}_{1}\,\left(c_{R}^{V,q'q}\right)^2 
-4\,Q_{5}\,\left(c_{L}^{V,q'q}\,c_{R}^{S,q'q}\right)
\right]\right\}\nonumber
\eea
In the final line we have used a Fierz identity to rearrange the 
operator $(V-A)\otimes(V+A)$ into the scalar operator $Q_5$.
Here ${\mathcal H}_{\rm eff}^{S}$ describes the effective Hamiltonian
for the case of scalar unparticles
and ${\mathcal H}_{\rm eff}^{V}$ for vector unparticles,
hence the total effective Hamiltonian is simply, 
\bea
{\mathcal H}_{\rm eff}^{q'q}={\mathcal H}_{\rm eff}^{S,q'q}
+{\mathcal H}_{\rm eff}^{V,q'q}
\eea
Above we have defined the quark operators $Q_1-Q_5$ as follows,
\bea
Q_1&=&{\ol{q}'}^{\alpha}_{\!\!L}\gamma_{\mu}q^{\alpha}_L\,\,
{\ol{q}'}^{\beta}_{\!\!L}\gamma^{\mu}q^{\beta}_L\\
Q_2&=&{\ol{q}'}^{\alpha}_{\!\!R}q^{\alpha}_L\,\,
{\ol{q}'}^{\beta}_{\!\!R}q^{\beta}_L\\
Q_3&=&{\ol{q}'}^{\alpha}_{\!\!R}q^{\beta}_L\,\,
{\ol{q}'}^{\beta}_{\!\!R}q^{\alpha}_L\\
Q_4&=&{\ol{q}'}^{\alpha}_{\!\!R}q^{\alpha}_L\,\,
{\ol{q}'}^{\beta}_{\!\!L}q^{\beta}_R\\
Q_5&=&{\ol{q}'}^{\alpha}_{\!\!R}q^{\beta}_L\,\,
{\ol{q}'}^{\beta}_{\!\!L}q^{\alpha}_R
\eea
Writing the effective Hamiltonian in terms of these operators we have,
\bea
{\mathcal H}_{\rm eff}^{q'q}
=\sum_{i=1}^5 (C^S_i+C^V_i) Q_i
+\sum_{i=1}^3 (\tilde{C}^S_i+\tilde{C}^V_i) \tilde{Q}_i
\eea
Here the operators 
$\tilde{Q}_{1,2,3}$ are obtained from $Q_{1,2,3}$ by the exchange
$L\leftrightarrow R$.

The hadronic matrix elements, taking into account for renormalization
effects, are defined as follows,
\bea
\langle\ol{M}^0 |Q_1(\mu)|M^0\rangle&=&\frac{1}{3}M_{M}f_{M}^2 B_1(\mu)\\
\langle\ol{M}^0 |Q_2(\mu)|M^0\rangle&=&-\frac{5}{24}M_{M}f_{M}^2 R(\mu)B_2(\mu)\\
\langle\ol{M}^0 |Q_3(\mu)|M^0\rangle&=&\frac{1}{24}M_{M}f_{M}^2 R(\mu) B_3(\mu)\\
\langle\ol{M}^0 |Q_4(\mu)|M^0\rangle&=&\frac{1}{4}M_{M}f_{M}^2 R(\mu)B_4(\mu)\\
\langle\ol{M}^0 |Q_5(\mu)|M^0\rangle&=&\frac{1}{12}M_{M}f_{M}^2 R(\mu)B_5(\mu)\\
{\rm where,}&R(\mu)&=\left(\frac{M_M}{m_{q}(\mu)+m_{q'}(\mu)}\right)^2
\eea

From Eq.~(\ref{Hscal}-\ref{Hvec}), it is straightforward to calculate
the Wilson coefficients for scalar unparticles which are as follows,
\bea
C^S_2&=& \frac{A_{d_\U}}{\sin d_\U\pi}\frac{e^{-i\phi_{\U}}}{M_M^4} 
\left(\frac{M_M^2}{\Lambda_\U^2}\right)^{d_\U}
\left(
m_{q'}c_L^{S,q'q}-m_{q}c_R^{S,q'q}
\right)
\left(
m_{q}c_R^{S,q'q}-m_{q'}c_L^{S,q'q}
\right)
\\
C^S_4&=& 2\frac{A_{d_\U}}{\sin d_\U\pi}\frac{e^{-i\phi_{\U}}}{M_M^4} 
\left(\frac{M_M^2}{\Lambda_\U^2}\right)^{d_\U}
\left(
m_{q'}c_L^{S,q'q}-m_{q}c_R^{S,q'q}
\right)
\left(
m_{q}c_L^{S,q'q}-m_{q'}c_R^{S,q'q}
\right)
\\
\tilde{C}^S_2&=& \frac{A_{d_\U}}{\sin d_\U\pi}\frac{e^{-i\phi_{\U}}}{M_M^4} 
\left(\frac{M_M^2}{\Lambda_\U^2}\right)^{d_\U}
\left(
m_{q'}c_R^{S,q'q}-m_{q}c_L^{S,q'q}
\right)
\left(
m_{q}c_L^{S,q'q}-m_{q'}c_R^{S,q'q}
\right)
\\
C^S_1&=&C^S_3=C^S_5=\tilde{C}^S_1=\tilde{C}^S_3=0
\eea
and for vector unparticles we find the following Wilson coefficients,
\bea
C^V_1&=&-\frac{A_{d_\U}}{\sin d_\U\pi}\frac{e^{-i\phi_{\U}}}{M_M^2} 
\left(\frac{M_M^2}{\Lambda_\U^2}\right)^{d_\U-1}
\left(c_L^{V,q'q}\right)^2
\\
C^V_2&=& \frac{A_{d_\U}}{\sin d_\U\pi}\frac{e^{-i\phi_{\U}}}{M_M^4} 
\left(\frac{M_M^2}{\Lambda_\U^2}\right)^{d_\U-1}
\left(
m_{q'}\,c_L^{V,q'q}-m_q \,c_R^{V,q'q}
\right)
\left(
m_q \,c_R^{V,q'q}-m_{q'}\,c_L^{V,q'q}
\right)
\\
C^V_4&=& 2\frac{A_{d_\U}}{\sin d_\U\pi}\frac{e^{-i\phi_{\U}}}{M_M^4} 
\left(\frac{M_M^2}{\Lambda_\U^2}\right)^{d_\U-1}
\left(
m_{q'} \,c_L^{V,q'q}-m_q \,c_R^{V,q'q}
\right)
\left(
m_{q} \,c_L^{V,q'q}-m_{q'} \,c_R^{V,q'q}
\right)
\\
C^V_5&=&4\frac{A_{d_\U}}{\sin d_\U\pi}\frac{e^{-i\phi_{\U}}}{M_M^2} 
\left(\frac{M_M^2}{\Lambda_\U^2}\right)^{d_\U-1}
\left(c_L^{V,q'q}c_R^{V,q'q}\right)\\
\tilde{C}^V_1&=&-\frac{A_{d_\U}}{\sin d_\U\pi}\frac{e^{-i\phi_{\U}}}{M_M^2} 
\left(\frac{M_M^2}{\Lambda_\U^2}\right)^{d_\U-1}
\left(c_R^{V,q'q}\right)^2
\\
\tilde{C}^V_2&=& \frac{A_{d_\U}}{\sin d_\U\pi}\frac{e^{-i\phi_{\U}}}{M_M^4} 
\left(\frac{M_M^2}{\Lambda_\U^2}\right)^{d_\U-1}
\left(
m_{q'}\,c_R^{V,q'q}-m_q \,c_L^{V,q'q}
\right)
\left(
m_q \,c_L^{V,q'q}-m_{q'}\,c_R^{V,q'q}
\right)
\\
C^V_3&=&\tilde{C}^V_3=0
\eea
where we have approximated $t=s\sim M_M^2$. From these two sets
of Wilson coefficients it is clear that the case of vector unparticles
not only includes more contributions as $C_1^V\neq 0$, $C_5^V\neq 0$, but
also that their Wilson coefficients are enhanced by a factor 
$\Lambda_{\U} / M_M^2$ compared to the scalar unparticle case. As a result the 
vector unparticle parameter space shall be suppressed by the same 
factor.

These Wilson coefficients will mix with each other as a result of 
renormalization group(RG) running down to the scale of $M_M$.
For the $B$ system, with a scale of new physics $\Lambda_\U=1$ TeV, 
these Wilson
coefficients at the scale $\mu_b=m_b$ may be approximated as 
\cite{Becirevic:2001jj},
\bea
C_1(\mu_b)&\approx& \,\,\,\,0.805 \, C_1(\Lambda_\U)\\
C_2(\mu_b)&\approx& \,\,\,\,1.988 \, C_2(\Lambda_\U)-0.417 \,C_3(\Lambda_\U)\\
C_3(\mu_b)&\approx& -0.024 \, C_2(\Lambda_\U)+0.496 \,C_3(\Lambda_\U)\\
C_4(\mu_b)&\approx& \,\,\,\,3.095 \, C_4(\Lambda_\U)+0.725 \,C_5(\Lambda_\U)\\
C_5(\mu_b)&\approx& \,\,\,\,0.086 \, C_4(\Lambda_\U)+0.884 \,C_5(\Lambda_\U)
\eea

The $\Delta F=2$ transitions are defined as,
\bea
\langle \ol{M}^0 |\mathcal{H}^{\Delta F=2}_{\rm eff}|M^0 \rangle
=M_{12}
\eea
with the meson mass eigenstate difference defined as,
\bea
\Delta M \equiv M_H - M_L = 2|M_{12}|
\eea

We can define in a model independent way the contribution to meson
mixings in the presence of new physics (NP) as,
\bea
M_{12}=M_{12}^{\rm SM}(1+R)
\label{totalM}
\eea
where $M_{12}^{\rm SM}$ denotes the SM contribution and 
$R\equiv r\,e^{i\,\sigma}=M_{12}^{\rm NP}/M_{12}^{\rm SM}$ 
parameterizes the NP contribution.

The associated CP phase may then be defined as,
\bea
\phi\equiv arg(M_{12})=\phi^{\rm SM}+\phi^{\rm NP}
\eea 
where $\phi^{\rm SM}=arg(M_{12}^{\rm SM})$ 
and $\phi^{\rm NP}=arg(1+r\,e^{i\,\sigma})$.

At this point it is important to make a further remark regarding 
the unparticle like model derived 
from a scale invariant theory of continuous mass fields
as discussed in \cite{Deshpande:2008ra}. Although this model 
contains no fixed point or dimensional transmutation and 
therefore does not correspond to 
unparticles as defined in \cite{hep-ph/0703260}, 
it does contain unparticle-like local operators coupling to the SM.
In this model it can be shown that the scale invariance will be broken by
interactions with SM fields, resulting in a mass gap which could
be rather large \cite{Fox:2007sy}.
In this setting we may then expect that there are unparticle states
lying below the scale of 1 TeV. Therefore the above approach of 
integrating out unparticle effects at the scale of 1 TeV would not be valid
in this special case, rather the Wilson coefficients should be instead
directly calculated at the low-energy scale. Such a calculation would offer
its own challenges, in particular the $t$-channel exchange would 
involve a nonlocal hadronic matrix element.
In this work we shall not consider such a case 
and instead assume that our unparticles
do not derive from a scale invariant theory of continuous mass fields.
Rather, we follow closely along the general framework 
set out by Georgi's original 
work in \cite{hep-ph/0703260} where there is dimensional transmutation 
at the scale $\Lambda_\U$.

\subsection{$B_{d,s}$ mixing and unparticle physics}

In this work we shall focus on the constraints imposed on 
unparticle physics couplings from $B_{d,s}$ mixing. Therefore
we set $q'=b$, $q=s,\,d$ and $M^0=B^0_s,\,B^0_d$.

In the $B$ system, the standard model contribution to $M_{12}^q$ is given by,
\be
M_{12}^{q,{\rm SM}}=\frac{G_F^2 M_W^2}{12 \pi^2}
M_{B_q}\hat{\eta}^B\,f_{B_q}^2 \hat{B}_{B_q}
(V_{tq}^* V_{tb})^2\, S_0(x_t)
\ee
where $G_F$ is Fermi's constant, $M_W$ the mass of the $W$ boson,
$\hat{\eta}^B=0.551$ \cite{Buras:1990fn} 
is a short-distance QCD correction identical 
for both the $B_s$ and $B_d$ systems. The bag parameter $\hat{B}_{B_q}$ 
and decay constant $f_{B_q}$ are nonperturbative quantities and contain
the majority of the theoretical uncertainty. $V_{tq}$ and $V_{tb}$ are elements
of the Cabibbo-Kobayashi-Maskawa (CKM) matrix 
\cite{Cabibbo:1963yz,Kobayashi:1973fv}, 
and $S_0(x_t\equiv \bar{m}_t^2/M_W^2)=2.34\pm 0.03$, 
with $\bar{m}_t(m_t)=164.5\pm 1.1$ GeV \cite{:2008vn}, is one of the 
Inami-Lim functions \cite{Inami:1980fz}.

We can now constrain both the magnitude and phase of the NP
contribution, $r_q$ and $\sigma_q$, through the comparison of 
the experimental measurements with SM expectations. From the 
definition of Eq.~(\ref{totalM}) we have the constraint,
\bea
\rho_q\equiv \frac{\Delta M_q}{\Delta M_q^{\rm SM}}
=\sqrt{1+2r_q\cos\sigma_q + r_q^2}
\label{rho}
\eea
The values for $\rho_q$ given by the UTfit analysis 
\cite{arXiv:0803.0659,arXiv:0707.0636} at the $95\%$ C.L. are,
\bea
\rho_d&=&\left[0.53,\,2.05\right]\\
\rho_s&=&\left[0.62,\,1.93\right]
\eea
These constraints on $\rho_q$ encode the CP conserving measurements
of $\Delta M_{d,s}$.
The phase associated with NP can also be 
written in terms of $r_q$ and $\sigma_q$,
\bea
\sin\phi_q^{\rm NP}&=&\frac{r_q\sin\sigma_q}
{\sqrt{1+2 r_q\cos\sigma_q+r_q^2}},
\nonumber\\
\cos\phi_q^{\rm NP}&=&\frac{1+r_q\cos\sigma_q}
{\sqrt{1+2 r_q\cos\sigma_q+r_q^2}}
\label{phiNP}
\eea
Here \cite{arXiv:0803.0659,arXiv:0707.0636} gives the $95\%$ C.L. constraints,
\bea
\phi_d^{\rm NP}&\hspace{-1mm}=&\hspace{-1mm}
\left[-16.6,\,3.2\right]^{\rm o}\label{phidNP}\\
\phi_{s}^{\rm NP}&\hspace{-1mm}=&\hspace{-1mm}
\left[-156.90,\,-106.40\right]^{\rm o}\cup
\left[-60.9,\,-18.58\right]^{\rm o}
\label{phisNP}
\eea
these constraint represent those of the CP phase measurements of $\phi_{d,s}$.

In order to consistently apply the above constraints all input 
parameters are chosen to match those used in the analysis
of the UTfit group \cite{arXiv:0803.0659,arXiv:0707.0636} with the 
nonperturbative parameters,
\bea
f_{B_s}\sqrt{\hat{B}_{B_s}}&=&262\pm 35 \,{\rm MeV}\\
\xi&=&1.23\pm 0.06 \,{\rm MeV}\\
f_{B_s}&=&230\pm 30 \, {\rm MeV}\\
f_{B_d}&=&189\pm 27 \, {\rm MeV}
\eea

\section{Numerical Analysis}\label{sec3}

In our analysis we shall first 
consider the case of scalar and vector unparticles
separately and further subdivide each into four classes as follows,
\begin{itemize}
\item { One real coupling}; $c_L\neq 0$ and $c_R=0$, with $c_L \in \mathbb{R}$
\item { Two real couplings}; $c_L\neq 0$ and $c_R\neq 0$, with $\{c_L,c_R\}\in\mathbb{R}$
\item { One complex coupling}; $c_L\neq 0$ and $c_R=0$, with $c_L\in\mathbb{C}$
\item { Two complex couplings}; $c_L\neq 0$ and $c_R\neq 0$, with $\{c_L,c_R\}\in\mathbb{C}$
\end{itemize}
Hence we have a total of eight scenarios to consider 
in the following analysis. 

For our analysis we shall take the unparticle
scale $\Lambda_\U=1$ TeV. In the literature \cite{arXiv:0705.1821,arXiv:0704.3532,arXiv:0705.0689,arXiv:0707.1234,arXiv:0707.1535,arXiv:0710.3663,arXiv:0711.3516,arXiv:0801.0895} it has been common to perform an analysis with a 
fixed value of the scaling dimension, $d_\U=\frac{3}{2}$. 
It was recently shown \cite{Grinstein:2008qk,Nakayama:2007qu}
that from unitarity considerations \cite{Mack:1975je} there exists 
a bound on the scaling dimension, $d_\U\geq j_1+j_2+2-\delta_{j_1 j_2,0}$.
Here $(j_1,j_1)$ are the operators Lorentz spins. In particular this leads 
to the bounds $d_\U\geq 2(3)$ for our scalar(vector) unparticle operators,
which conflicts with the choice $\frac{3}{2}$.
First it should be mentioned that these bounds are derived from 
a conformal field theory(CFT) point of view. 
Here we are considering a scale invariant 
sector at high energies, and in general scale invariant theories need not 
also be invariant under the conformal group. The specific example of 
Banks-Zaks fields though, is in
fact invariant under the full conformal group 
\cite{Grinstein:2008qk,Nakayama:2007qu}.

 \begin{center}
 \begin{figure}[h]
 \includegraphics[width=6.5truecm,clip=true]{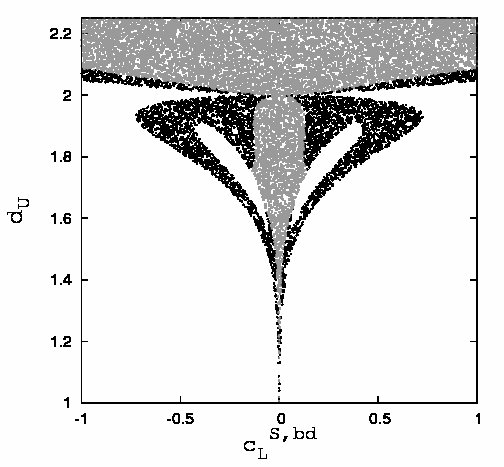}
 \includegraphics[width=6.5truecm,clip=true]{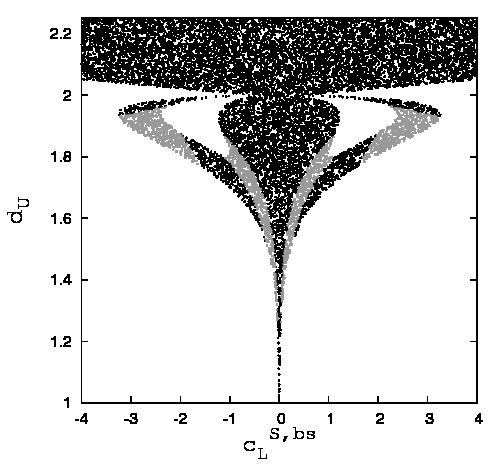}
 \caption{Constraints on the $d_\U$ versus $c_L^S$ parameter space 
from $B_d$ mixing(left) and $B_s$ mixing(right) for the case of 
a single real coupling $c_L^S\neq 0$ and $c_R^S=0$. Black points indicate the 
$\Delta M_{d,s}$ allowed regions, while grey points indicate the regions
are in agreement with both $\Delta M_{q}$ and the CP phase $\phi_q$.}
\label{fig2}
 \end{figure}
\vskip-5mm
 \end{center}
In order to make a connection with the literature, and to quantify the 
impact of the recent measurement of the CP phase $\phi_s$, we choose to fix
the scaling dimension at $d_\U=\frac{3}{2}$. Through simple scaling
of the coefficient $S_{d_\U}$, it is also possible to 
make a connection to a value of the 
scaling dimension in the region suggested by the above
mentioned unitarity bound. 
In general the bounds on unparticle couplings 
with the scaling dimension in the region $d_\U\geq 2(3)$
shall be much weaker than for $d_\U=\frac{3}{2}$, unless the value is 
close to an integer, where there is a pole in the unparticle propagator.
For example, if we write the scaling dimension for 
scalar unparticles as $d_\U=2+\epsilon_S$, with $\epsilon_S$ a small positive 
number, then we have $|S_{d_\U=3/2}|=|S_{d_\U=2+\epsilon_S}|$ for a value
of $\epsilon_S\approx 0.00021$. Hence the bounds derived here from the 
CP conserving quantities $\Delta M_{d,s}$ shall also apply to the case 
$d_\U\approx 2.00021$. The situation is not so straightforward for the 
bounds derived from CP violating constraints of $\phi_{d,s}$. 
For vector unparticles we have a similar situation for $d_\U=3+\epsilon_V$, 
with $\epsilon_V\approx 2\times 10^{-11}$.
The above unitarity bound shall be considered in 
more detail elsewhere \cite{next:unpart}.

\subsection{Scalar unparticles}

In this section we shall analyze the contribution to $B_{d,s}$ mixing
from scalar unparticles taking a number of example scenarios.
In each case we shall use the measurements of $\Delta M_{d,s}$
as well as the CP phases $\phi_{d,s}$ to constrain the unparticle 
parameter space.

\subsubsection{One real coupling: $c_L^S\neq 0$, $c_R^S=0$\label{case1}}

In this case the Wilson coefficients at the sale $\Lambda_\U$ simplify to,
\bea
C^S_2&=& -\frac{A_{d_\U}}{\sin d_\U\pi}\frac{e^{-i\phi_{\U}}}{M_M^4} 
\left(\frac{M_M^2}{\Lambda_\U^2}\right)^{d_\U}
m_{b}^2\,\left(c_L^{S,bq}\right)^2
\\
C^S_4&=& 2\frac{A_{d_\U}}{\sin d_\U\pi}\frac{e^{-i\phi_{\U}}}{M_M^4} 
\left(\frac{M_M^2}{\Lambda_\U^2}\right)^{d_\U}
m_{b}\,m_{q}\,\left(c_L^{S,bq}\right)^2
\\
\tilde{C}^S_2&=& -\frac{A_{d_\U}}{\sin d_\U\pi}\frac{e^{-i\phi_{\U}}}{M_M^4} 
\left(\frac{M_M^2}{\Lambda_\U^2}\right)^{d_\U}
m_{q}^2\,\left(c_L^{S,bq}\right)^2
\label{case1WC}
\\
C^S_1&=&C^S_3=C^S_5=\tilde{C}^S_1=\tilde{C}^S_3=0
\eea
where $q=d,s$ correspond to $B_{d}$ or $B_s$ mixing. 

In this first case, in order to have two free parameters,
we shall allow the scaling dimension $d_\U$ to vary. In general the 
unparticle effects shall be smaller for larger $d_\U$. 
In this case the scaling dimension $d_\U$ also uniquely
determines the NP CP phase 
and as such the small CP phase 
allowed by $\phi_d$ and the large CP phase required by the recent 
measurement of $\phi_s$ will act as a stringent constraint on 
the $d_\U-c_L^S$ parameter space. 

Fig.~\ref{fig2}
displays the constraint on the $d_\U-c_L^S$ parameter space from the 
$B_d$ mixing(left panel) and $B_s$ mixing(right panel), including CP phase
constraints. As expected the general feature of these plots is that for 
small $d_\U$ the mixing contribution from unparticles is large and so the 
couplings $c_L^{S,bq}$ are strongly constrained. For larger values of $d_\U$
the opposite is true with $d_\U>2$ resulting in no bound on $c_L^{S,bq}$ from 
either $\Delta M_d$ or $\Delta M_s$. 
 \begin{center}
 \begin{figure}[h]
 \includegraphics[width=6.5truecm,clip=false]{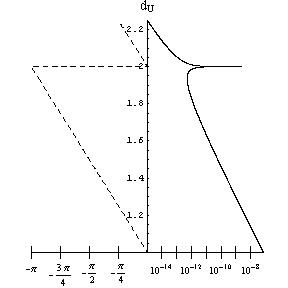}
 \caption{Plot of the variation of the phase and magnitude of the 
quantity $S_{d_\U}$ with the scaling dimension $d_\U$.}
\label{fig3}
 \end{figure}
\vskip-5mm
 \end{center}

In the case of scalar unparticles 
the Wilson coefficients have the common quantity, 
\bea
S_{d_\U}&=&\frac{A_{d_\U}}{\sin d_\U\pi}\frac{e^{-i\phi_{\U}}}{M_M^4} 
\left(\frac{M_M^2}{\Lambda_\U^2}\right)^{d_\U}
\eea
Here the NP CP phase is determined solely by $\arg(S_{d_\U})$
such that $\sigma_q=\arg(S_{d_\U}/M_{12}^{\rm SM})$.
The magnitude of $S_{d_\U}$ also determines the size of the 
unparticle contribution to $B$ mixing and hence determines the level
of constraint on the coupling $c_L^S$.
Fig.~\ref{fig3} shows the variation with $d_\U$ of the phase and magnitude 
of $S_{d_\U}$, from which we can see that the 
phase of $S_{d_\U}$ is constrained to be in the range $(-\pi,0)$.
The magnitude of $S_{d_\U}$ is generally decreasing for increasing $d_\U$
except for when $d_\U$ approaches $2$ where there is a singularity due to 
the factor $1/\sin d_\U\pi$. The general features of Fig.~\ref{fig2} are now 
easy to understand. The large value of $S_{d_\U}$ for small $d_\U$
restricts the allowed values of $c_L^S$ to a very narrow region. 
For example, for $d_\U=\frac{3}{2}$ we have the bounds,
\bea
|c_L^{S,bd}|&<&0.067\\
|c_L^{S,bs}|&<&0.23
\eea
The difference in the level of the above constraints on the couplings 
$c_L^{S,bd}$ and $c_L^{S,bs}$ are due to the ratio of CKM matrix 
elements $V_{td}/V_{ts}\approx \lambda$.
Increasing $d_\U$ causes $S_{d_\U}$ to decrease sharply, resulting in an 
increase in the allowed size of the couplings $c_L^S$ as seen in 
Fig.~\ref{fig2}. 
Approaching $d_\U=2$ causes a rapid increase 
in $S_{d_\U}$ and thus the allowed values
of $c_L^S$ are again heavily constrained. For $d_\U>2$ the quantity
$S_{d_\U}$ becomes increasingly small, resulting in no bound on the 
couplings $c_L^S$.
 \begin{center}
 \begin{figure}[h]
 \includegraphics[width=6.5truecm,clip=true]{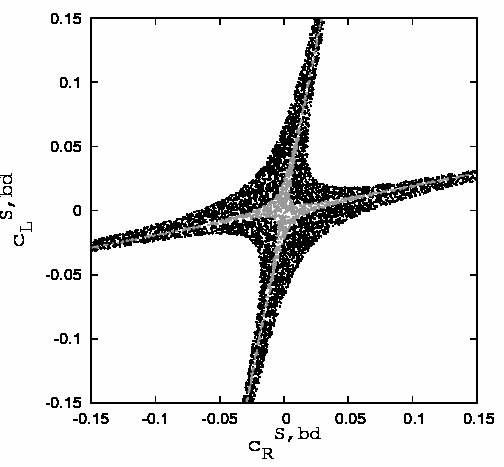}
 \includegraphics[width=6.5truecm,clip=true]{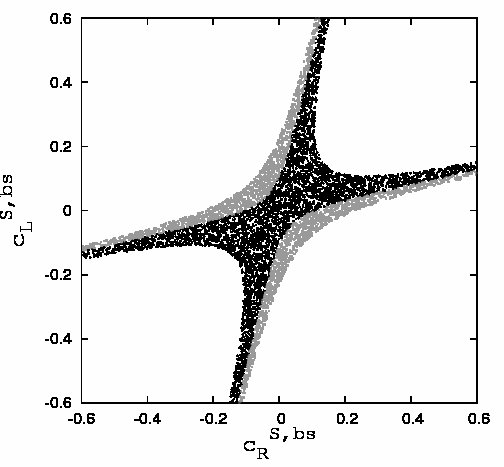}
 \caption{ Plot of the allowed $c_L^S$-$c_R^S$ parameter space 
 for $B_d$ mixing(left) and $B_s$ mixing(right) 
in the case of two real couplings. 
Black points show regions
which agree with the measurement of $\Delta M_{d,s}$ while grey points show
additional agreement with the measurement of the CP phases $\phi_{d,s}$.}
\label{fig4}
 \end{figure}
\vskip-5mm
 \end{center}

We can see from Fig.~\ref{fig2} that there are windows in the allowed 
parameter space of $d_\U-c_L^S$.
From Eq.~(\ref{rho}) we can determine $r_q$ in terms of $\sigma_q$ as,
\bea
r_q=-\cos\sigma_q\pm\sqrt{\rho_q^2-\sin^2\sigma_q}
\label{rq}
\eea
and for $\rho_q<1$ the phase $\sigma_q$ is constrained to be in the region,
\bea
-\pi-\arcsin \rho_q \leq \sigma_q \leq -\pi+\arcsin \rho_q
\label{rhobound}
\eea
Taking the minimum allowed value of $\rho_q$ we can determine these
windows as corresponding to,
\bea
-\pi-\arcsin \rho_{q}^{min}+\phi_q^{\rm SM} 
\leq \arg(S_{d_\U}) 
\leq -\pi+\arcsin \rho_q^{min}+\phi_q^{\rm SM} 
\label{window}
\eea
with $\phi_s^{\rm SM}=-2\beta_s=-0.0409$ 
and $\phi_d^{\rm SM}=2\beta=0.781$, determines these windows to be,
\bea
{\rm B_d:}\hspace{0.5cm} 1.56<&d_\U&<1.92\\
{\rm B_s:}\hspace{0.5cm} 1.80<&d_\U&<2.00 
\eea
To the inside of these windows, with smaller values of the coupling $c_L^S$,
the unparticle contribution is small enough to satisfy the mixing 
constraint. To the outside of these windows, with larger values of $c_L^S$,
the unparticle contribution is larger than the SM contribution. 
When the phase $\sigma_q$ is in the range $(-\pi,-\frac{\pi}{2})$
the quantity $\cos\sigma_q$ in Eq.~(\ref{rho}) is negative, allowing larger 
values of $r_q$. This solution corresponds to the case when the 
unparticle contribution carries the opposite sign to the SM and 
turns over the sign of $M_{12}$.

 \begin{center}
 \begin{figure}[h]
 \includegraphics[width=6.5truecm,clip=true]{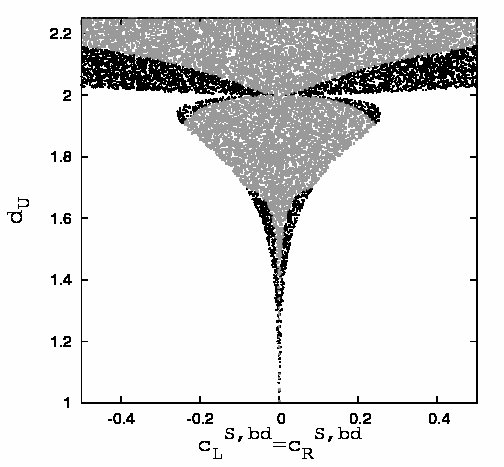}
 \includegraphics[width=6.5truecm,clip=true]{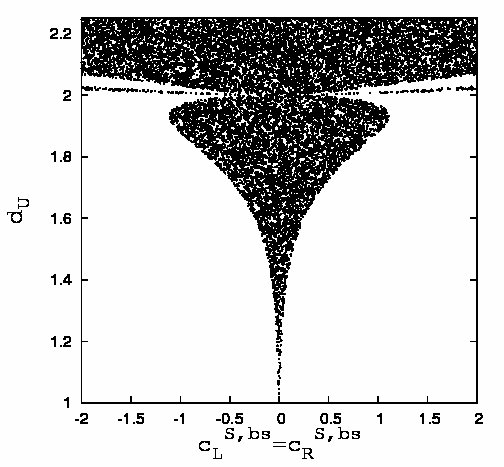}
 \caption{Variation of allowed parameter space 
of the real coupling $c_L^S=c_R^S$ 
with scaling dimension $d_\U$ for $B_d$ mixing (left) and $B_s$ mixing (right).
Black plotted points agree with the CP conserving mixing quantities
$\Delta M_{d,s}$, while grey points also agree with the CP phases 
$\phi_{d,s}$.}
\label{fig5} 
\end{figure}
\vskip-5mm
 \end{center}
The black points of Fig.~\ref{fig2} correspond to the region of parameter
space allowed by the measurement of the $B$ mixing parameters $\Delta M_{d,s}$,
while grey points also satisfy the constraint from the measurement of 
the CP phases $\phi_{d,s}$ respectively. The left panel of Fig.~\ref{fig2}
shows that the allowed region is strongly constrained by
the  measurement of $\phi_d$ in addition to the constraint of $\Delta M_d$.
This additional constraint restricts the parameter space of $c_L^{s,bd}$
considerably. For example, for $d_\U=\frac{3}{2}$ and including the 
CP constraint we have the improved bound,
\bea
|c_L^{S,bd}|&<&0.024
\eea
which is almost three times smaller than the bound from 
$\Delta M_d$ alone.
As shown in Eq.~(\ref{phisNP}), in the $B_s$ system there 
is still a two fold ambiguity in the measurement
of the CP phase $\phi_s$. This two fold ambiguity is then seen
in the right panel of Fig.~\ref{fig2} as two distinct pairs of 
grey regions. 
The pair of grey regions with larger $c_L^{S,bs}$, and 
consequently larger $r_s$, corresponds to 
$\phi_s^{\rm NP}=\left[-156.90,\,-106.40\right]^{\rm o}$. This 
region is clearly disfavoured as it corresponds to the coupling
$c_L^{S,bs}$ being outside of the perturbative region. 
On the other hand the pair of grey
regions with smaller $c_L^{S,bs}$ and $r_s$, 
corresponds to
$\phi_s^{\rm NP}=\left[-60.9,\,-18.58\right]^{\rm o}$.
In this case the coupling $c_L^S$ is constrained to be in a narrow
band away from zero. For example with $d_\U=\frac{3}{2}$ we have a two-sided
bound, 
\bea
0.106 < |c_L^{S,bs}| < 0.23
\eea
The small $c_L^{S,bs}$ region also indicates that a value of the 
scaling dimension in the range $1.22 < d_\U < 1.87$ is preferred. 

Despite the two fold ambiguity in $\phi_s$ 
it is clear that values of the scaling dimension in the range $d_\U<2$
are preferred by the measurement of $\phi_s$.
Combining these $\phi_s$ preferred values for $d_\U$ 
with the constraints from $\phi_d$ implies a general bound on $c_L^{S,bd}$
as,
\bea
|c_L^{S,bd}| < 0.13
\eea 

\subsubsection{Two real coupling: $c_L^S\neq 0$, $c_R^S\neq 0$\label{case2}}

In this second case we shall allow the left and right unparticle 
couplings $c_L^S$ and $c_R^S$ to both be real and nonzero.
For this analysis we shall fix the scaling dimension $d_\U=\frac{3}{2}$.
The allowed $c_L^S-c_R^S$ parameter space is plotted in Fig.~\ref{fig4}
with black points constrained by $\Delta M_{d,s}$ and grey points further
constrained by the CP phase $\phi_{d,s}$. The allowed region extends out
along two lines where there is a cancellation between $c_L^S$ and $c_R^S$.
As a result no direct bound can be set on the unparticle couplings.

Starting from the definition of $R_q=M_{12}^{q,\rm NP}/M_{12}^{q,\rm SM}$ 
and taking the limit $m_q\rightarrow 0$ we arrive at an approximate relation,
\bea
6\left[(c_L^{S,bq})^2+(c_R^{S,bq})^2\right]
-32\,c_L^{S,bq}c_R^{S,bq}\approx r_q\,\epsilon_q
\eea
where the small quantity $\epsilon_q$ is defined as, 
\bea
\epsilon_q=
\frac{|M_{12}^{q,\rm SM}|}{M_{B_q} f_{B_q}^2 |S_{d_\U}|}
\eea
Here it is clear that a large cancellation between the couplings 
$c_L^S$ and $c_R^S$ is possible.
From Eq.~(\ref{rq}) we find that $(\rho_q-1)^2\leq r_q^2 \leq (\rho_q+1)^2$
which leads to,
\bea
 -0.1\,x_q \leq r_q\,\epsilon_q \leq x_q
\eea
with $x_s=1$ and $x_d=\lambda^2$.
The solution to these equations is then
a parabola in the $c_L^S-c_R^S$ parameter space
described by, 
\bea
c_L^{S,bq}=\frac{16}{6}c_R^{S,bq}
\pm\frac{1}{6}\sqrt{220\left(c_R^{S,bq}\right)^2+6\,\epsilon_q}
\eea
If we set $\epsilon_q\approx 0$, then these allowed regions follow 
along the lines,
\bea
c_L^{S,bq}\approx 5\,c_R^{S,bq},\hspace{0.5cm}
c_L^{S,bq}\approx \frac{1}{5}\,c_R^{S,bq}
\label{lines}
\eea
in good agreement with the results of the full calculation shown in 
Fig.~\ref{fig4}.
 \begin{center}
 \begin{figure}[h]
 \includegraphics[width=6.5truecm,clip=true]{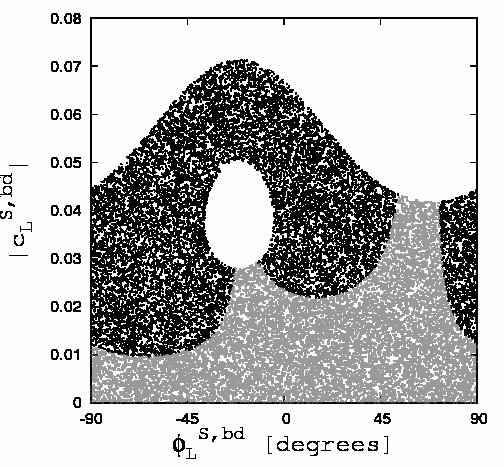}
 \includegraphics[width=6.5truecm,clip=true]{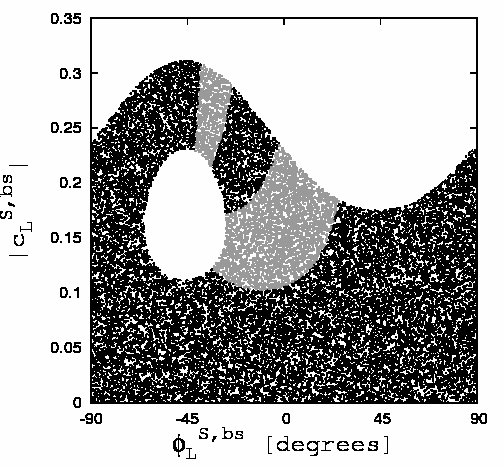}
 \caption{ Plot of the allowed $|c_L^S|$-$\phi_L^S$ parameter space 
 for $B_d$ mixing (left) and $B_s$ mixing (right) 
for the case of a single complex coupling
$c_L^S\neq 0$ and $c_R^S=0$.
Black plotted points agree with the CP conserving mixing quantities
$\Delta M_{d,s}$, while grey points also agree with $\phi_{d,s}$.}
\label{fig6}
 \end{figure}
\vskip-5mm
 \end{center}

In this case it is possible to have very large values of the couplings, 
as long as they lie approximately on the lines shown in Eq.~(\ref{lines}).
These solutions with large cancellations are clearly highly fine-tuned.
If we consider the case of $c_L^S=c_R^S$ then we have no such 
fine-tuning and we can extract the bounds,
\bea
c_L^{S,bd}\equiv c_R^{S,bd}&<& 2.5\times 10^{-2}\\
c_L^{S,bs}\equiv c_R^{S,bs}&<& 0.14
\eea
from the constraints of $\rho_{d,s}$.
From the grey regions of Fig.~\ref{fig4} we again see that the measurement
of the CP phases $\phi_{d,s}$ further constrains the allowed parameter space.
In the $B_d$ system the allowed region is reduced by more than a half.
For the case $c_L^S=c_R^S$ the CP constraint provides a much improved bound,
\bea
c_L^{S,bd}\equiv c_R^{S,bd}&<& 6.7\times 10^{-3}
\eea
In the $B_s$ system the new CP measurement 
of $\phi_s$ prefers points away from the 
origin. It is also interesting to notice that the case,
 $c_L^{S,bs}=c_R^{S,bs}$ is disfavoured by the latest measurement of 
$\phi_s$ at the $3\sigma$ level.

The case of equal left and right couplings is shown in more detail in 
Fig.~\ref{fig5} where the allowed parameter space is plotted as a 
function of the scaling dimension $d_\U$. Fig.~\ref{fig5} shows the allowed
parameter space for both $B_d$ (left panel) and $B_s$ (right panel).
From the left panel we can see that the allowed parameter space is rather 
similar to that shown in Fig.~\ref{fig2}. This time the windows in the 
plots have disappeared for the case of $B_d$ and have moved in the case of 
$B_s$. When we have $c_L^S=c_R^S$ we get a new 
contribution from $C_4^S$ not present in the case of Sec.~\ref{case1}
with $c_R^S=0$. The introduction of $C_4^S$ changes the overall sign of 
$M_{12}^{NP}$ and so in this case these holes obey a modified 
relation similar to Eq.~(\ref{window}) except there is no factor of $-\pi$.
This means that for the case of $B_d$ there is no such window and for 
$B_s$ the window is now at $2<d_\U<2.23$, as shown in Fig.~\ref{fig5}.
In this case the impact of the CP phase measurement $\phi_d$
is to again reduce the allowed parameter space. The impact of the recently
measured $\phi_s$ is much more profound, as can be seen from the right panel of
Fig.~\ref{fig5} where there are no grey points and only black regions.
The lack of any grey regions in this plot indicates that this case 
cannot satisfy the $3\sigma$ measurement of $\phi_s$ for any choice of 
the scaling dimension $d_\U$.
 \begin{center}
 \begin{figure}[h]
 \includegraphics[width=6.5truecm,clip=true]{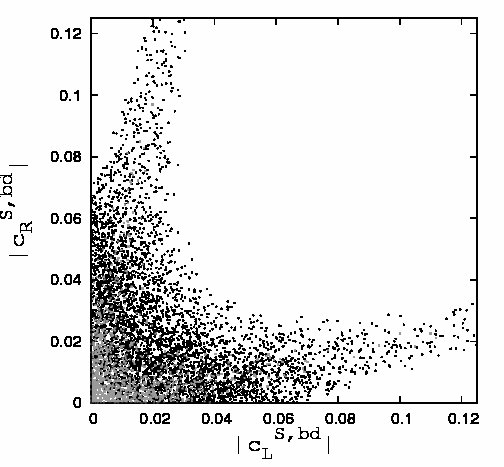}
 \includegraphics[width=6.5truecm,clip=true]{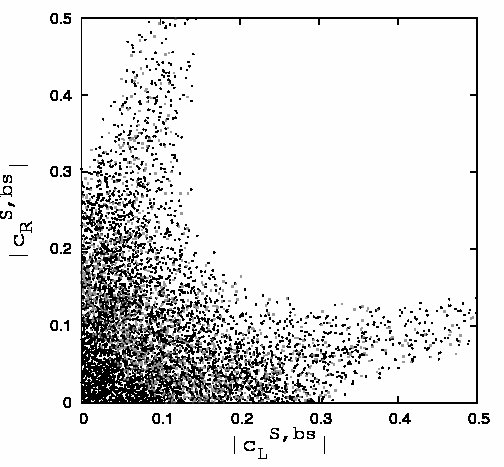}
 \caption{ Plot of the allowed $|c_L^S|$-$|c_R^S|$ parameter space 
 for $B_d$ mixing(left) and $B_s$ mixing(right) 
for the case of two complex couplings
$c_L^S\neq 0$ and $c_R^S\neq 0$.
Black plotted points agree with the CP conserving quantities
$\Delta M_{d,s}$, while grey points also agree with $\phi_{d,s}$.}
\label{fig7} 
\end{figure}
\vskip-5mm
 \end{center}

\subsubsection{One complex coupling: $c_L^S\neq 0$, $c_R^S= 0$\label{case3}}

In this case we have $c_R^S=0$ and one complex coupling $c_L^S$, with the 
scaling dimension also fixed at $d_\U=\frac{3}{2}$. In this case the 
Wilson coefficients again take the simplified form of Eq.~(\ref{case1WC}),
this time with a complex $c_L^{S,bq}$. As a result the phase of 
$M_{12}^{\rm NP}$ is not only determined by the scaling dimension, but also
by the phase of the coupling $c_L^{S,bq}$. 

Due to the hierarchy $m_b \gg m_q$ the dominant contribution here comes from
the Wilson coefficient $C_2^S$. 
For the purpose of simplicity we can make an approximation of the 
unparticle contribution as,
\bea
R_q=\frac{M_{12}^{q,\rm NP}}{M_{12}^{q,\rm SM}}
=\delta\,\left(c_L^{S,bq}\right)^2
\eea
where, 
$\delta\approx\frac{5}{24}m_b^2 M_{B_q}f_{Bq}^2 B_1 S_{d_\U}/M_{12}^{\rm SM}$,
with $|\delta|\approx 9.45$. This then leads to,
\bea
r_q&=& |\delta|\,\left|c_L^{S,bq}\right|^2\\
\sigma_q&=&
\arg\left(\left(c_L^{S,bq}\right)^2 S_{d_\U}/M_{12}^{q,\rm SM}\right),
\eea
For $d_{\U}=\frac{3}{2}$, we have 
$\sigma_q=2\phi_L^{S,bq}-\frac{\pi}{2}-\phi_q^{\rm SM}$, where
$\phi_L^{S,bq}=\arg\left(c_L^{S,bq}\right)$.
From Eq.~(\ref{rho}) we then have,
\bea
\rho_q=\sqrt{1+
2|\delta|\left|c_L^{S,bq}\right|^2
\cos(2\phi_L^{S,bq}-\phi_q^{\rm SM}-\frac{\pi}{2})
+|\delta|^2 \left|c_L^{S,bq}\right|^4}
\eea
The solutions to this equation are shown in Fig.~\ref{fig6} plotted
in the $|c_L^S|-\phi^S_L$ plane. Again we can see from Eq.~(\ref{rhobound})
that there are holes in the parameter space corresponding to,
$-38.63^{\rm o}<\phi_L^{S,bd}<-6.62^{\rm o}$ and $-65.33<\phi_L^{S,bs}<-27.01$,
as determined by Eq.~(\ref{rhobound}).
Using only the black points of Fig.~\ref{fig6} we can extract a bound 
on the magnitude of the couplings as follows,
\bea
|c_L^{S,bd}|&<&  7.1\times 10^{-2} \\
|c_L^{S,bs}|&<&  0.31
\eea
 \begin{center}
 \begin{figure}[h]
 \includegraphics[width=6.5truecm,clip=true]{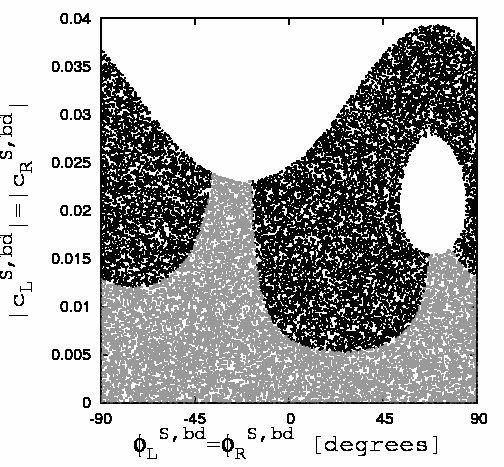}
 \includegraphics[width=6.5truecm,clip=true]{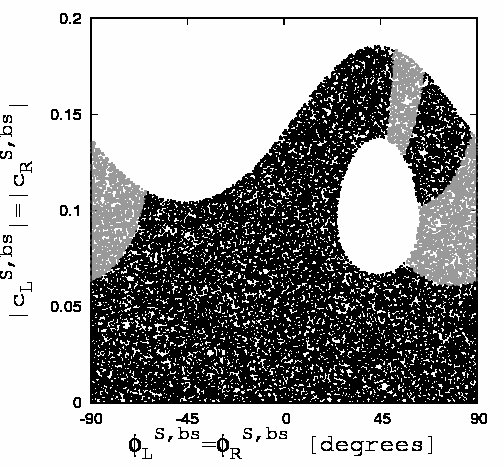}
 \caption{ Plot of the allowed $|c_L^S|$-$\phi_L^S$ parameter space 
 for $B_d$ mixing(left) and $B_s$ mixing(right) 
for the case of one complex coupling $c_L^S=c_R^S$.
Black plotted points agree with the CP conserving mixing quantities
$\Delta M_{d,s}$, while grey points also agree with $\phi_{d,s}$.}
\label{fig8} 
\end{figure}
\vskip-5mm
 \end{center}

From the grey regions plotted in Fig.~\ref{fig6} 
we can again see that the inclusion of the CP phase constraints leads to a
much reduced parameter space and improved bounds on the unparticle couplings,
\bea
&|c_L^{S,bd}|&<  4.3\times 10^{-2} \\
0.10<&|c_L^{S,bs}|&<  0.31
\eea
In addition we may also constrain the phase of the unparticle coupling
$c_L^{V,bs}$ in the case of $B_s$ mixing as,
\bea
-41.77^{\rm o}<\phi_L^{S,bs}<25.55^{\rm o}
\eea

\subsubsection{Two complex couplings: $c_L^S\neq 0$, $c_R^S\neq 0$\label{case4}}

In this final scalar unparticle case we take two 
complex couplings, $c_L^{S,bq}$ and $c_R^{S,bq}$, with the scale
dimension fix at $d_\U=\frac{3}{2}$.
Fig.~\ref{fig7} displays the allowed parameter space for this scenario.
In a similar way to the case with both couplings real, see Fig.~\ref{fig4},
the allowed regions extend along the lines 
$|c_L^{S,bq}|\approx 5\,|c_R^{S,bq}|$
and $|c_L^{S,bq}|\approx \frac{1}{5}\,|c_R^{S,bq}|$. 
No general bound can be placed
on the size of the unparticle couplings as large cancellations may occur 
even for very large values of the couplings. These solutions with large 
cancellations are however
very highly tuned and unattractive. Taking the subset of this case 
with the 
left and right-handed couplings equal, we get the bound,
\bea
|c_L^{S,bd}|=|c_R^{S,bd}|&<&0.039 \\
|c_L^{S,bs}|=|c_R^{S,bs}|&<&0.19
\eea
In this case the additional constraints from the CP phase measurements
of $\phi_{d,s}$ have very little effect on the allowed parameter space.
From the left panel of Fig.~\ref{fig7} it is clear from the clustering 
of grey points near the origin that the CP phase $\phi_d$ prefer smaller
values of $|c_L^{S,bd}|$. On the other hand from the right panel we see that
the opposite is true for $|c_L^{S,bs}|$ where there are only 
black points near the origin.

In Fig.~\ref{fig8} we study in more detail 
the case of equal and complex couplings
$c_L^S=c_R^S$. Fig.~\ref{fig8} shows the 
experimentally allowed $|c_L^S|$-$\phi_L^S$ parameter
space plotted for scaling dimension $d_\U=\frac{3}{2}$.
 \begin{center}
 \begin{figure}[h]
 \includegraphics[width=6.5truecm,clip=true]{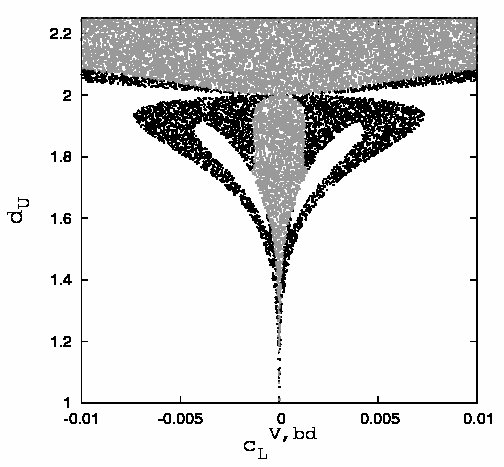}
 \includegraphics[width=6.5truecm,clip=true]{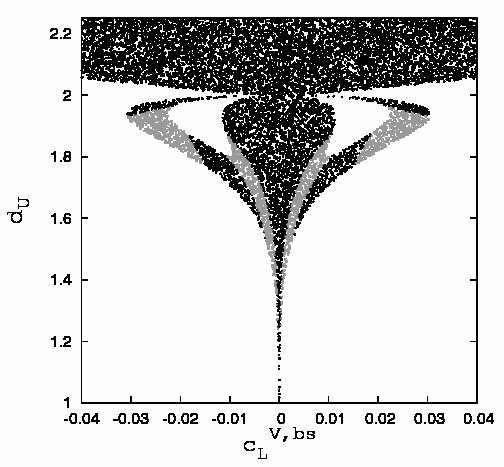}
 \caption{ Plot of the variation of the allowed $c_L^V$ parameter space
with the scaling dimension $d_\U$ 
 for $B_d$ mixing(left) and $B_s$ mixing(right) 
in the case of a single real coupling
$c_L^V\neq 0$ and $c_R^V=0$.
Black plotted points agree with the CP conserving mixing quantities
$\Delta M_{d,s}$, while grey points also agree with $\phi_{d,s}$.}
\label{fig9}
 \end{figure}
\vskip-5mm
 \end{center}

The left and right panels of Fig.~\ref{fig8} display exactly the features
just described. The grey allowed regions in the left panel corresponding to 
points satisfying both $\Delta M_d$ and $\phi_d$ constraints prefer lower
values of the coupling in the range,
\bea
|c_L^{S,bd}|=|c_R^{S,bd}|&<&0.024
\eea
From the right panel we can see that the phase $\phi_s$ indeed prefers 
regions with a larger coupling in the region,
\bea
0.06<&|c_L^{S,bs}|=|c_R^{S,bs}|&<0.18
\eea
In Fig.~\ref{fig8} we again see holes in the $|c_L^S|-\phi_L^S$ parameter
space. In the case of $c_L^S=c_R^S$  we have 
$\sigma_q=2\phi_L^{S,bq}+\pi+\arg(S_{d_\U})-\phi^{\rm SM}_q$, where the 
extra $\pi$ is from a change in sign of $M_{12}^{\rm NP}$ relative to 
the case in Sec.~\ref{case3}. Using Eq.~(\ref{rhobound}) we find that
these holes correspond to,
\bea
24.67^{\rm o}<&\phi_L^{S,bd}&<62.99^{\rm o}\\
51.37^{\rm o}<&\phi_L^{S,bs}&<83.38^{\rm o}
\eea

\subsection{Vector unparticles}

Now let us consider the case of vector unparticles. 
In general the contributions from vector unparticles to 
meson mixing is larger than 
from scalar unparticles due to the enhancement of the 
Wilson coefficients by the factor 
$\Lambda^2_{\U} / M_{B_q}^2$. As a result 
the experimentally allowed parameter space for vector unparticles 
will in general also be suppressed by this same factor.

\subsubsection{One real coupling: $c_L^V\neq 0$, $c_R^V= 0$\label{case5}}

First we shall consider the case of a single real 
vector unparticle coupling, $c_L^{V,bq}$. Assuming $c_R^{V,bq}=0$ we 
then choose to allow the scaling dimension $d_\U$ to vary and study the
experimentally allowed parameter space. This case is rather similar
to that of scalar unparticle case discussed in Sec.~\ref{case1}.
The allowed regions of parameter space for this case are shown in 
Fig.~\ref{fig9} with black points indicating the regions of parameter
space allowed by the measurement 
of $\Delta M_{d,s}$, while grey points indicate 
in addition agreement with measurements of the CP phase $\phi_{d,s}$.
Fig.~\ref{fig9} again shows that 
the measurement of the CP phases $\phi_{d,s}$ improves
greatly on the constraints set by $\Delta M_{d,s}$. 
For example with $d_\U=\frac{3}{2}$ the constraint from 
$\Delta M_{d,s}$ alone gives, 
\bea
|c_L^{V,bd}|&<&6.8\times 10^{-4} \\
|c_L^{V,bs}|&<&2.2\times 10^{-3}
\eea
Adding the restriction from the measurement of the CP phases $\phi_{d,s}$
we get the improved bounds,
\bea
&|c_L^{V,bd}|&< 2.4\times 10^{-4} \\
1\times 10^{-3}<&|c_L^{V,bs}|&< 2.2\times 10^{-3}
\eea
 \begin{center}
 \begin{figure}[h]
 \includegraphics[width=6.5truecm,clip=true]{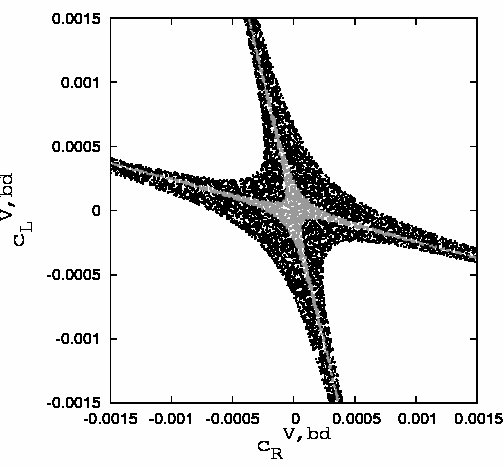}
 \includegraphics[width=6.5truecm,clip=true]{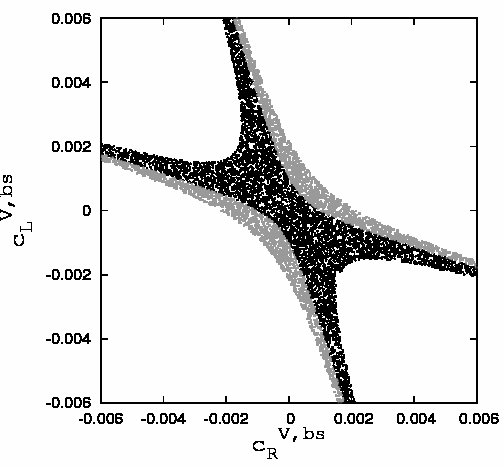}
 \caption{Plot of the allowed $c_L^V$-$c_R^V$ parameter space 
 for $B_d$ mixing(left) and $B_s$ mixing(right) 
in the case of two real couplings
$c_L^V\neq 0$ and $c_R^V\neq 0$.
Black plotted points agree with the CP conserving mixing quantities
$\Delta M_{d,s}$, while grey points also agree with $\phi_{d,s}$.}
\label{fig10}
 \end{figure}
\vskip-5mm
 \end{center}

Looking at the grey allowed regions of the right panel of Fig.~\ref{fig9}
 it is clear that the recently 
measured phase in $B_s$ mixing, $\phi_s$, prefers values of the scaling
dimension in the range $1.22<d_\U<1.96$. With the scaling dimension $d_\U$
in the range $1<d_\U<2$ we may extract the general bounds,
\bea
&|c_L^{V,bd}|&< 1.3\times 10^{-3} \\
2.4\times 10^{-4}<&|c_L^{V,bs}|&< 3.1 \times 10^{-2}
\eea

\subsubsection{Two real couplings: $c_L^V\neq 0$, $c_R^V\neq 0$\label{case6}}

In this case we fix the scaling dimension $d_\U=\frac{3}{2}$ with
two real nonzero couplings $c_L^{V,bq}$ and $c_R^{V,bq}$.
Fig.~\ref{fig10} shows the allowed regions of the $c_L^{V,bq}$-$c_R^{V,bq}$
parameter space.
Here the unparticle contribution to the $\Delta F=2$ effective
Hamiltonian satisfies the approximate relation,
\bea
2\left[(c_L^{V,bq})^2+(c_R^{V,bq})^2\right]
+12\,c_L^{S,bq}c_R^{S,bq}\approx r_q\,\epsilon_q
\frac{M_{B_q}^2}{\Lambda_\U^2}
\eea
where the small quantity $\epsilon_q$ is as defined in section~\ref{case2}.
Again solutions extend along two lines of the approximate form,
\bea
c_L^{V,bq}&\approx& -0.3\,c_R^{V,bq}\\
c_L^{V,bq}&\approx& -3.3\,c_R^{V,bq}
\eea
as can be seen in Fig.~\ref{fig10}. In general no bound can be placed on 
the size of the unparticle coupling in this case
due to the possibility of large cancellations. 
Let us then again consider the 
simplified case with left and right unparticle couplings set equal to 
each other, $c_L^{V,bq}=c_R^{V,bq}$.
First using the constraints from $\Delta M_{d,s}$ 
we may set the bounds,
\bea
|c_L^{V,bd}|=|c_R^{V,bd}|&<&2.7\times 10^{-4}\\
|c_L^{V,bs}|=|c_R^{V,bs}|&<&9.2\times 10^{-4}
\eea
The grey plotted points of Fig.~\ref{fig10} show the effect of the additional
constraint from the CP phase measurements of $\phi_{d,s}$.
The allowed parameter space is once again much reduced by the 
inclusion of these CP constraints.
The CP phase $\phi_d$ improves the above bound
on $c_L^{V,bd}$ to,
\bea
|c_L^{V,bd}|=|c_R^{V,bd}|&<&9.6\times 10^{-5}.
\eea
while the CP phase constraint of $\phi_s$ improves the above bound
on $c_L^{V,bs}$ to,
\bea
4.2\times 10^{-4}<&|c_L^{V,bd}|=|c_R^{V,bd}|&<9.2\times 10^{-4}.
\eea
when $d_\U=\frac{3}{2}$.
 \begin{center}
 \begin{figure}[h]
 \includegraphics[width=6.5truecm,clip=true]{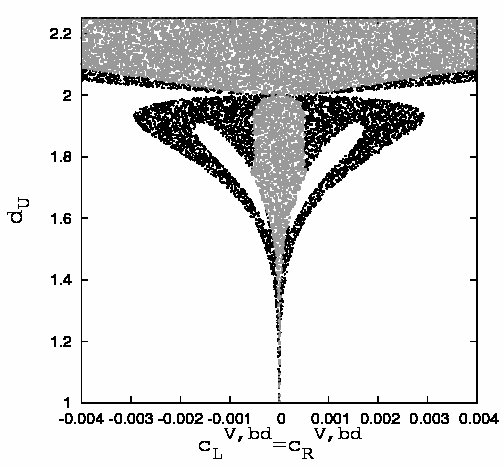}
 \includegraphics[width=6.5truecm,clip=true]{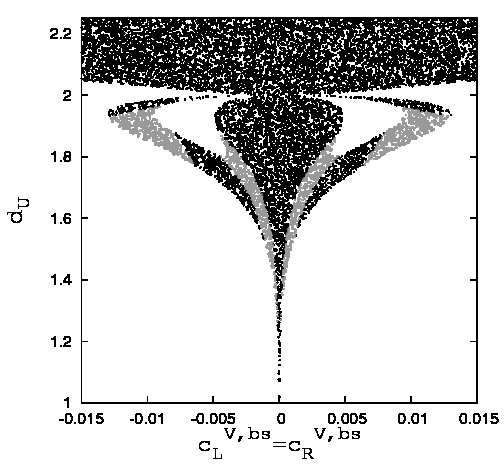}
 \caption{Plot of the variation of the allowed parameter space with the 
scaling dimension $d_\U$
 for $B_d$ mixing(left) and $B_s$ mixing(right) in
the case of one real coupling $c_L^V=c_R^V$.
Black plotted points agree with the CP conserving mixing quantities
$\Delta M_{d,s}$, while grey points also agree with $\phi_{d,s}$.}
\label{fig11} 
\end{figure}
\vskip-5mm
 \end{center}

The case of $c_L^V=c_R^V$ is displayed in Fig.~\ref{fig11} which 
very much resembles the case of section~\ref{case5} with just one 
real coupling $c_L^V\neq 0$ and $c_R^V=0$. Here with two identical
real couplings $c_L^V=c_R^V$ the allowed parameter space is
less than half that found in section~\ref{case5}.
As such all bounds on the couplings 
found in section~\ref{case1} may be restated
for this case by simply multiplying by $0.42$ for $B_s$ and $0.4$ for $B_d$.
 \begin{center}
 \begin{figure}[h]
 \includegraphics[width=6.5truecm,clip=true]{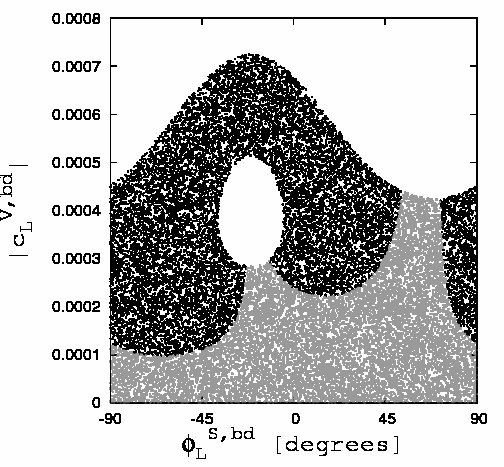}
 \includegraphics[width=6.5truecm,clip=true]{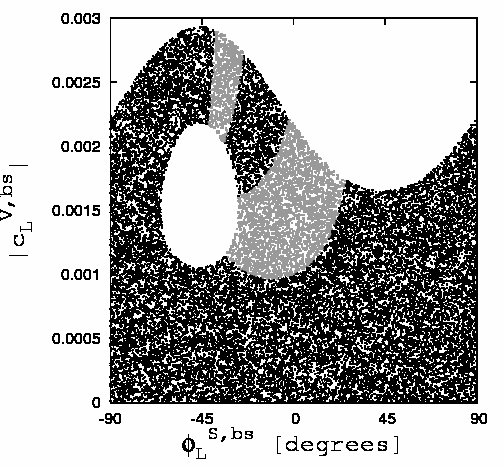}
 \caption{ Plot of the allowed $|c_L^V|$-$\phi_L^V$ parameter space 
 for $B_d$ mixing(left) and $B_s$ mixing(right) 
in the case of one complex coupling
$c_L^V\neq 0$, $c_R^V=0$.
Black plotted points agree with the CP conserving mixing quantities
$\Delta M_{d,s}$, while grey points also agree with $\phi_{d,s}$.}
\label{fig12}
 \end{figure}
\vskip-5mm
 \end{center}

\subsubsection{One complex coupling: $c_L^V\neq 0$, $c_R^V= 0$\label{case7}}

Next we consider the case of a 
single complex coupling and fix the scaling dimension
to be $d_\U=\frac{3}{2}$. This case is similar to that analyzed in 
section~\ref{case3}. The allowed parameter space for this case is shown in
Fig.~\ref{fig12}. 
The black points in Fig.~\ref{fig12} show the regions of parameter space
which agree with experimental measurements of $\Delta M_{d,s}$ while 
grey points indicate the additional agreement with the CP phases $\phi_{d,s}$.
From the black points we can extract the bound,
\bea
|c_L^{S,bd}|&<&  7.2\times 10^{-4} \\
|c_L^{S,bs}|&<&  2.9\times 10^{-3}
\eea

It is clear from the grey plotted points of Fig.~\ref{fig12} that
the CP phase constraints have an important role to play in this 
scenario also. From these grey regions we may again extract improved
bounds in the magnitude of the couplings,
\bea
&|c_L^{S,bd}|&<  4.4\times 10^{-4}\\
9.7 \times 10^{-4}<&|c_L^{S,bs}|&<  2.9\times 10^{-3}
\eea
In addition we may also constrain the phase of the unparticle coupling
$c_L^{V,bs}$ as,
\bea
-41.80^{\rm o}<\phi_L^{V,bs}<25.52^{\rm o}
\eea
 \begin{center}
 \begin{figure}[h]
 \includegraphics[width=6.5truecm,clip=true]{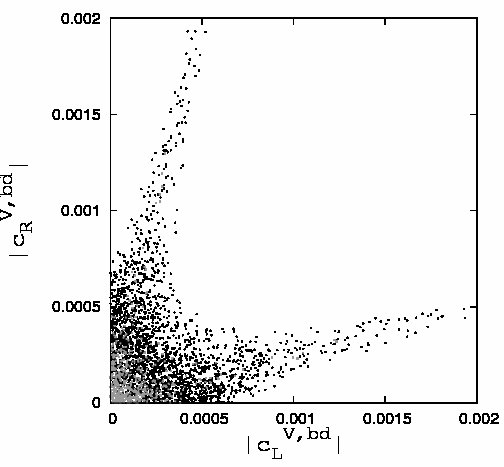}
 \includegraphics[width=6.5truecm,clip=true]{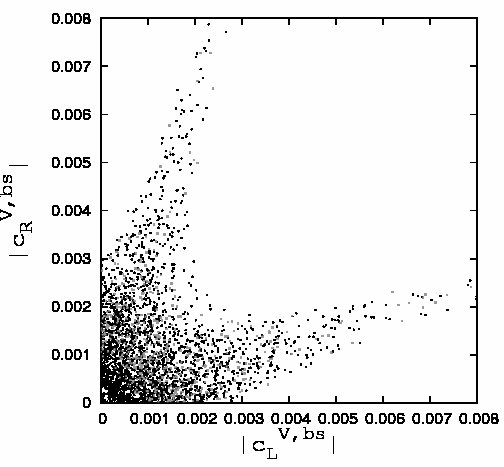}
 \caption{ Plot of the allowed $|c_L^V|$-$|c_R^V|$ parameter space 
 for $B_d$ mixing(left) and $B_s$ mixing(right) 
for the case of two complex couplings.
Black plotted points agree with the CP conserving mixing quantities
$\Delta M_{d,s}$, while grey points also agree with $\phi_{d,s}$.}
\label{fig13}
 \end{figure}
\vskip-5mm
 \end{center}

\subsubsection{Two Complex Couplings: $c_L^V\neq 0$, $c_R^V\neq 0$\label{case8}}

In this final scenario we take the general case of two independent 
complex couplings $c_L^{V,bq}$ and $c_R^{V,bq}$, and fix $d_\U=\frac{3}{2}$.
The experimentally allowed parameter space for $q=d,s$ are shown in
Fig.~\ref{fig13}, again showing the allowed regions from both $\Delta M_{d,s}$
and $\phi_{d,s}$ constraints. This case is closely related to that of 
section~\ref{case4} where we found that the allowed regions followed along 
two lines. Again here the allowed regions approximately 
follow the lines $|c_L^{V,bq}|=0.3\,|c_R^{V,bq}|$ and 
$|c_L^{V,bq}|=\,3.3\,|c_R^{V,bq}|$ and 
as a result no general bound can be set in this case either.
From the grey regions plotted in
Fig.~\ref{fig13} we can see that the CP phases $\phi_{d,s}$
prefer smaller $|c^{V,bd}|$ and larger $|c^{V,bs}|$.

Let us consider 
the case when the two couplings are equal. This case is plotted in 
Fig.~\ref{fig14} in the plane of $|c_L^V|$-$\phi_L^V$.
The CP conserving $B_{d,s}$ mixing constraints shown by the black 
plotted points may provide the bounds,
\bea
|c_L^{S,bd}|&<& 2.9\times 10^{-4} \\
|c_L^{S,bs}|&<& 1.2\times 10^{-3}
\eea
When we include the CP phase constraints of $\phi_{d,s}$ we get the 
improved limits,
\bea
&|c_L^{S,bd}|&< 1.8\times 10^{-4} \\
4\times 10^{-4}<&|c_L^{S,bs}|&< 1.2\times 10^{-3}
\eea
In this case we may again constrain the phase of the unparticle coupling
$c_L^{V,bs}$ as,
\bea
-41.80^{\rm o}<\phi_L^{V,bs}<25.52^{\rm o}
\eea
 \begin{center}
 \begin{figure}[h]
 \includegraphics[width=6.5truecm,clip=true]{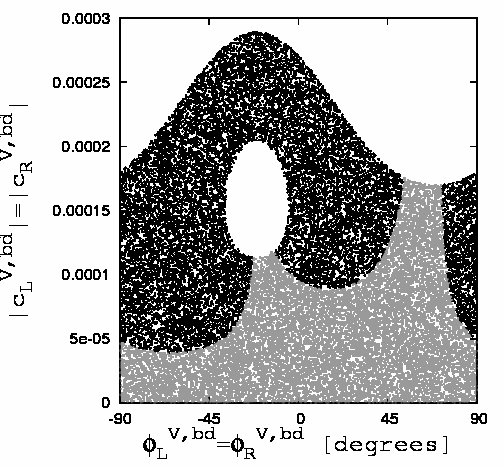}
 \includegraphics[width=6.5truecm,clip=true]{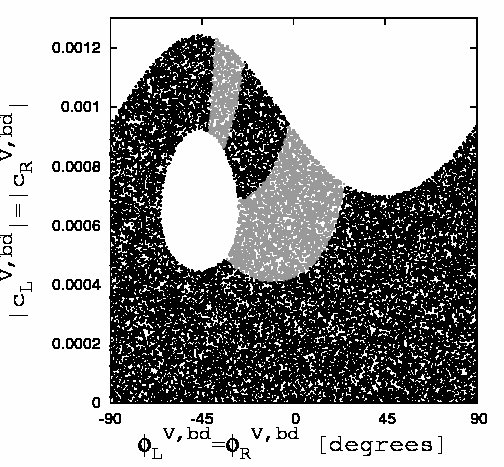}
 \caption{ Plot of the allowed $|c_L^V|$-$\phi_L^V$ parameter space 
 for $B_d$ mixing(left) and $B_s$ mixing(right) 
for the case of one complex coupling 
$c_L^V=c_R^V$.
Black plotted points agree with the CP conserving mixing quantities
$\Delta M_{d,s}$, while grey points also agree with $\phi_{d,s}$.}
\label{fig14} 
\end{figure}
\vskip-5mm
 \end{center}

 \section{Conclusions}\label{conc}

In this work we have studied the contribution of the recently 
suggested unparticle physics to $B$ meson mixing. Our main aim 
was to study the impact of the CP phase associated with $B_{d,s}$
mixing. In particular we have been interested in the impact of the
recent $3\sigma$ evidence for new physics
found in the phase of $B_s$ mixing. The phase of $B_d$
mixing is rather close to the standard model expectation and as such
these two phases provide very different constraints on the allowed 
phase of new physics.

For our numerical analysis we have chosen to study the cases of 
scalar unparticles and vector unparticles separately. Each case has 
further been subdivided into a number of scenarios depending on the 
type  unparticle couplings. In each scenario we have been careful to 
understand and discuss the specific characteristics of the allowed
parameter space and, in particular, to understand the impact of the CP
phase constraints on this parameter space and, 
where possible, to set bounds on the
unparticle couplings. Our main goal here has been to identify the impact
of the phase constraints on the allowed parameter space in each scenario
and, in particular, that of the recently 
measured phase of $B_s$ mixing, $\phi_s$.
A summary of the bounds placed on the unparticle couplings from 
the mixing parameters $\Delta M_{d,s}$ alone and for $\Delta M_{d,s}$ together
with $\phi_{d,s}$ are shown in Table~\ref{table1}.

\begin{table}[h]
\begin{center}
\begin{tabular}{|c||c|c||c|c|}
\hline
Scalar unparticles & $\Delta M_{d}$ & $\Delta M_{s}$ & $\Delta M_{d}$ \& $\phi_{d}$  & $\Delta M_{s}$ \& $\phi_{s}$\\  
\hline
$c_L^S\in \mathbb{R}$, $c_R^S=0$ & 
$|c_L^{S,bd}|<0.067$ &$|c_L^{S,bs}|<0.23$ &$|c_L^{S,bd}|<0.024$ &$0.106<|c_L^{S,bs}|<0.23$ \\
$c_L^S,\,c_R^S\in \mathbb{R}$ & 
no constraint &no constraint &no constraint &no constraint  \\
$c_L^S=c_R^S\in \mathbb{R}$ & 
$|c_L^{S,bd}|<0.025$ & $|c_L^{S,bs}|<0.14$ &$|c_L^{S,bd}|<0.0067$ & Excluded \\
$c_L^S\in \mathbb{C}$, $c_R^S=0$ & 
$|c_L^{S,bd}|<0.071$ & $|c_L^{S,bs}|<0.31$ &$|c_L^{S,bd}|<0.043$ &$0.10<|c_L^{S,bs}|<0.31$ \\
$c_L^S,\,c_R^S\in \mathbb{C}$ & 
no constraint &no constraint &no constraint &no constraint  \\
$c_L^S=c_R^S\in \mathbb{C}$ & 
$|c_L^{S,bd}|<0.039$ & $|c_L^{S,bs}|<0.19$ &$|c_L^{S,bd}|<0.024$ &$0.06<|c_L^{S,bs}|<0.18$ \\
\hline
\hline
Vector unparticles & $\Delta M_{d}$ & $\Delta M_{s}$ & $\Delta M_{d}$ \& $\phi_{d}$  & $\Delta M_{s}$ \& $\phi_{s}$\\  
\hline
$c_L^V\in \mathbb{R}$, $c_R^V=0$ & 
$|c_L^{S,bd}|<6.8\times 10^{-4}$ &$|c_L^{S,bs}|<2.2\times 10^{-3}$ &$|c_L^{S,bd}|<2.4\times 10^{-4}$ &$1\times 10^{-3}<|c_L^{S,bs}|<2.2\times 10^{-3}$ \\
$c_L^V,\,c_R^V\in \mathbb{R}$ & 
no constraint &no constraint &no constraint &no constraint  \\
$c_L^V=c_R^V\in \mathbb{R}$ & 
$|c_L^{S,bd}|<2.7\times 10^{-4}$ & $|c_L^{S,bs}|<9.2\times 10^{-4}$ &$|c_L^{S,bd}|<9.6\times 10^{-5}$ & $4.2\times 10^{-4}<|c_L^{S,bs}|<9.2\times 10^{-4}$ \\
$c_L^V\in \mathbb{C}$, $c_R^V=0$ & 
$|c_L^{S,bd}|<7.2\times 10^{-4}$ & $|c_L^{S,bs}|<2.9\times 10^{-3}$ &$|c_L^{S,bd}|<4.4\times 10^{-4}$ &$9.7\times 10^{-4}<|c_L^{S,bs}|<2.9\times 10^{-3}$ \\
$c_L^V,\,c_R^V\in \mathbb{C}$ & 
no constraint &no constraint &no constraint &no constraint  \\
$c_L^V=c_R^V\in \mathbb{C}$ & 
$|c_L^{S,bd}|<2.9\times 10^{-4}$ & $|c_L^{S,bs}|<1.2\times 10^{-3}$ &$|c_L^{S,bd}|<1.8\times 10^{-4}$ &$4\times 10^{-4}<|c_L^{S,bs}|<1.2 \times 10^{-3}$ \\
\hline
\end{tabular}
\caption
{
Summary of bounds on unparticle couplings with scaling dimension
$d_\U=\frac{3}{2}$ from $95\%$ C.L.  measurement of 
$\Delta M_{d,s}$ alone and also from a combination
of the $95\%$ C.L. measurements of 
both $\Delta M_{d,s}$ and $\phi_{d,s}$.
}\label{table1}
\end{center}
\vskip-5mm
\end{table}

The main conclusion of this work is that 
the CP phase constraints do 
indeed improve markedly on the CP conserving 
bounds set on unparticle couplings
in every scenario studied by a factor of $2\sim 4$, see table~\ref{table1}. 
Of particular interest is the case of
scalar unparticles with real and equal couplings $c_L^{S,bs}=c_R^{S,bs}$
for the $B_s$ system. Using only the measurement of $\Delta M_s$ we have
a bound in the region of 0.14, but by adding the constraint
from the $95\%$ C.L. measurement of $\phi_s$ this case is excluded.


\begin{thebibliography}{99}
\bibitem{arXiv:0704.2457}
  H.~Georgi,
  Phys.\ Lett.\  B {\bf 650} (2007) 275
  [arXiv:0704.2457 [hep-ph]].

\bibitem{hep-ph/0703260}
  H.~Georgi,
  Phys.\ Rev.\ Lett.\  {\bf 98}, 221601 (2007)
  [arXiv:hep-ph/0703260].

\bibitem{Nucl.Phys.B196.189}
  T.~Banks and A.~Zaks,
  Nucl.\ Phys.\  B {\bf 196} (1982) 189.

\bibitem{arXiv:0704.2588}
  K.~Cheung, W.~Y.~Keung and T.~C.~Yuan,
  Phys.\ Rev.\ Lett.\  {\bf 99} (2007) 051803
  [arXiv:0704.2588 [hep-ph]].

\bibitem{arXiv:0704.3532}
  M.~Luo and G.~Zhu,
  Phys.\ Lett.\  B {\bf 659} (2008) 341
  [arXiv:0704.3532 [hep-ph]].

\bibitem{arXiv:0705.0794}
  G.~J.~Ding and M.~L.~Yan,
  Phys.\ Rev.\  D {\bf 76} (2007) 075005
  [arXiv:0705.0794 [hep-ph]].

\bibitem{arXiv:0705.2622}
  M.~Duraisamy,
  arXiv:0705.2622 [hep-ph].

\bibitem{arXiv:0705.3518}
  N.~Greiner,
  Phys.\ Lett.\  B {\bf 653} (2007) 75
  [arXiv:0705.3518 [hep-ph]].

\bibitem{arXiv:0705.4542}
  T.~M.~Aliev, A.~S.~Cornell and N.~Gaur,
  JHEP {\bf 0707} (2007) 072
  [arXiv:0705.4542 [hep-ph]].

\bibitem{arXiv:0705.4599}
  P.~Mathews and V.~Ravindran,
  Phys.\ Lett.\  B {\bf 657} (2007) 198
  [arXiv:0705.4599 [hep-ph]].

\bibitem{arXiv:0706.1284}
  Y.~Liao and J.~Y.~Liu,
  Phys.\ Rev.\ Lett.\  {\bf 99} (2007) 191804
  [arXiv:0706.1284 [hep-ph]].

\bibitem{arXiv:0706.3155}
  K.~Cheung, W.~Y.~Keung and T.~C.~Yuan,
  Phys.\ Rev.\  D {\bf 76} (2007) 055003
  [arXiv:0706.3155 [hep-ph]].

\bibitem{arXiv:0707.0187}
  S.~L.~Chen, X.~G.~He and H.~C.~Tsai,
  JHEP {\bf 0711} (2007) 010
  [arXiv:0707.0187 [hep-ph]].

\bibitem{arXiv:0707.0893}
  T.~Kikuchi and N.~Okada,
  Phys.\ Lett.\  B {\bf 661} (2008) 360
  [arXiv:0707.0893 [hep-ph]].

\bibitem{arXiv:0707.2074}
  D.~Choudhury and D.~K.~Ghosh,
  arXiv:0707.2074 [hep-ph].

\bibitem{arXiv:0708.0036}
  M.~Neubert,
  Phys.\ Lett.\  B {\bf 660} (2008) 592
  [arXiv:0708.0036 [hep-ph]].

\bibitem{arXiv:0708.0671}
  M.~x.~Luo, W.~Wu and G.~h.~Zhu,
  Phys.\ Lett.\  B {\bf 659} (2008) 349
  [arXiv:0708.0671 [hep-ph]].

\bibitem{arXiv:0708.3802}
  A.~T.~Alan and N.~K.~Pak,
  arXiv:0708.3802 [hep-ph].

\bibitem{arXiv:0708.1960}
  E.~L.~Berger, M.~M.~Block, D.~W.~McKay and C.~I.~Tan,
  Phys.\ Rev.\  D {\bf 77} (2008) 053007
  [arXiv:0708.1960 [hep-ph]].

\bibitem{arXiv:0709.2478}
  M.~C.~Kumar, P.~Mathews, V.~Ravindran and A.~Tripathi,
  arXiv:0709.2478 [hep-ph].

\bibitem{arXiv:0709.1505}
  S.~Mantry, M.~Trott and M.~B.~Wise,
  Phys.\ Rev.\  D {\bf 77} (2008) 013006
  [arXiv:0709.1505 [hep-ph]].

\bibitem{arXiv:0710.2230}
  K.~Cheung, W.~Y.~Keung and T.~C.~Yuan,
  arXiv:0710.2230 [hep-ph].

\bibitem{arXiv:0710.4239}
  A.~T.~Alan, N.~K.~Pak and A.~Senol,
  arXiv:0710.4239 [hep-ph].

\bibitem{arXiv:0710.5773}
  O.~Cakir and K.~O.~Ozansoy,
  arXiv:0710.5773 [hep-ph].

\bibitem{arXiv:0711.1665}
  I.~Sahin and B.~Sahin,
  arXiv:0711.1665 [hep-ph].

\bibitem{arXiv:0711.3361}
  K.~Cheung, C.~S.~Li and T.~C.~Yuan,
  arXiv:0711.3361 [hep-ph].

\bibitem{arXiv:0805.1150}
  T.~M.~Aliev, A.~Bekmezci and M.~Savci,
  arXiv:0805.1150 [hep-ph].

\bibitem{arXiv:0805.0199}
  X.~G.~He and C.~C.~Wen,
  arXiv:0805.0199 [hep-ph].

\bibitem{Mureika:2007nc}
  J.~R.~Mureika,
  Phys.\ Lett.\  B {\bf 660}, 561 (2008)
  [arXiv:0712.1786 [hep-ph]].

\bibitem{Alan:2007rg}
  A.~T.~Alan,
  arXiv:0711.3272 [hep-ph].

\bibitem{liaoy1}
Y.~Liao,
arXiv:0804.4033 [hep-ph].

\bibitem{liaoy2}
Y.~Liao, 
arXiv:0804.0752 [hep-ph].

\bibitem{liaoy3}
Y.~Liao, 
Eur.\ Phys.\ J.C55:483,2008. arXiv:0708.3327 [hep-ph].

\bibitem{liaoy4}
Y.~Liao, J.~Y.~Liu, Phys.\ Rev.\ Lett. \ 99:191804, 2007.
arXiv:0706.1284 [hep-ph].

\bibitem{liaoy5}
Y.~Liao, Phys.\ Rev.\ D76:056006, 2007.
arXiv:0705.0837 [hep-ph]

\bibitem{arXiv:0705.0689}
  C.~H.~Chen and C.~Q.~Geng,
  Phys.\ Rev.\  D {\bf 76} (2007) 115003
  [arXiv:0705.0689 [hep-ph]].

\bibitem{arXiv:0706.0850}
  C.~H.~Chen and C.~Q.~Geng,
  Phys.\ Rev.\  D {\bf 76} (2007) 036007
  [arXiv:0706.0850 [hep-ph]].

\bibitem{arXiv:0707.1268}
  C.~S.~Huang and X.~H.~Wu,
  arXiv:0707.1268 [hep-ph].

\bibitem{arXiv:0709.0235}
  C.~H.~Chen and C.~Q.~Geng,
  Phys.\ Lett.\  B {\bf 661} (2008) 118
  [arXiv:0709.0235 [hep-ph]].

\bibitem{Zwicky:2007yc}
  R.~Zwicky,
  J.\ Phys.\ Conf.\ Ser.\  {\bf 110} (2008) 072050
  [arXiv:0710.4430 [hep-ph]].

\bibitem{Zwicky:2007vv}
  R.~Zwicky,
  Phys.\ Rev.\  D {\bf 77} (2008) 036004
  [arXiv:0707.0677 [hep-ph]].

\bibitem{arXiv:0705.1821}
  X.~Q.~Li and Z.~T.~Wei,
  Phys.\ Lett.\  B {\bf 651} (2007) 380
  [arXiv:0705.1821 [hep-ph]].

\bibitem{arXiv:0707.1234}
  R.~Mohanta and A.~K.~Giri,
  Phys.\ Rev.\  D {\bf 76} (2007) 075015
  [arXiv:0707.1234 [hep-ph]].

\bibitem{arXiv:0707.1535}
  A.~Lenz,
  Phys.\ Rev.\  D {\bf 76} (2007) 065006
  [arXiv:0707.1535 [hep-ph]].

\bibitem{arXiv:0710.3663}
  S.~L.~Chen, X.~G.~He, X.~Q.~Li, H.~C.~Tsai and Z.~T.~Wei,
  arXiv:0710.3663 [hep-ph].

\bibitem{arXiv:0711.3516}
  R.~Mohanta and A.~K.~Giri,
  Phys.\ Lett.\  B {\bf 660} (2008) 376
  [arXiv:0711.3516 [hep-ph]].

\bibitem{arXiv:0801.0895}
  C.~H.~Chen, C.~S.~Kim and Y.~W.~Yoon,
  arXiv:0801.0895 [hep-ph].

\bibitem{arXiv:0705.1326}
  T.~M.~Aliev, A.~S.~Cornell and N.~Gaur,
  Phys.\ Lett.\  B {\bf 657} (2007) 77
  [arXiv:0705.1326 [hep-ph]].

\bibitem{arXiv:0705.2909}
  C.~D.~Lu, W.~Wang and Y.~M.~Wang,
  Phys.\ Rev.\  D {\bf 76} (2007) 077701
  [arXiv:0705.2909 [hep-ph]].

\bibitem{arXiv:0709.3435}
  G.~j.~Ding and M.~L.~Yan,
  arXiv:0709.3435 [hep-ph].

\bibitem{arXiv:0711.2744}
  E.~O.~Iltan,
  arXiv:0711.2744 [hep-ph].

\bibitem{arXiv:0806.2944}
  Z.~T.~Wei, Y.~Xu and X.~Q.~Li,
  arXiv:0806.2944 [hep-ph].

\bibitem{arXiv:0802.4015}
  A.~Hektor, Y.~Kajiyama and K.~Kannike,
  arXiv:0802.4015 [hep-ph].

\bibitem{arXiv:0802.1277}
  E.~O.~Iltan,
  arXiv:0802.1277 [hep-ph].

\bibitem{arXiv:0705.3636}
  H.~Davoudiasl,
  Phys.\ Rev.\ Lett.\  {\bf 99} (2007) 141301
  [arXiv:0705.3636 [hep-ph]].

\bibitem{arXiv:0708.1404}
  S.~Hannestad, G.~Raffelt and Y.~Y.~Y.~Wong,
  Phys.\ Rev.\  D {\bf 76} (2007) 121701
  [arXiv:0708.1404 [hep-ph]].

\bibitem{arXiv:0708.2812}
  P.~K.~Das,
  Phys.\ Rev.\  D {\bf 76} (2007) 123012
  [arXiv:0708.2812 [hep-ph]].

\bibitem{arXiv:0708.4339}
  A.~Freitas and D.~Wyler,
  JHEP {\bf 0712} (2007) 033
  [arXiv:0708.4339 [hep-ph]].

\bibitem{arXiv:0710.4275}
  G.~L.~Alberghi, A.~Y.~Kamenshchik, A.~Tronconi, G.~P.~Vacca and G.~Venturi,
  Phys.\ Lett.\  B {\bf 662} (2008) 66
  [arXiv:0710.4275 [hep-th]].

\bibitem{arXiv:0711.1506}
  T.~Kikuchi and N.~Okada,
  arXiv:0711.1506 [hep-ph].

\bibitem{arXiv:0803.3223}
  Y.~Gong and X.~Chen,
  arXiv:0803.3223 [astro-ph].

\bibitem{liaoy6}
S.~L.~Chen, X.~G.~He, X.~P.~Hu, Y.~Liao,
arXiv:0710.5129 [hep-ph].

\bibitem{arXiv:0706.0302}
  S.~Zhou,
  Phys.\ Lett.\  B {\bf 659} (2008) 336
  [arXiv:0706.0302 [hep-ph]].

\bibitem{arXiv:0706.0325}
  G.~J.~Ding and M.~L.~Yan,
  arXiv:0706.0325 [hep-ph].

\bibitem{arXiv:0707.2285}
  X.~Q.~Li, Y.~Liu and Z.~T.~Wei,
  arXiv:0707.2285 [hep-ph].

\bibitem{arXiv:0708.3485}
  D.~Majumdar,
  arXiv:0708.3485 [hep-ph].

\bibitem{arXiv:0709.0678}
  L.~Anchordoqui and H.~Goldberg,
  Phys.\ Lett.\  B {\bf 659} (2008) 345
  [arXiv:0709.0678 [hep-ph]].

\bibitem{arXiv:0707.2132}
  H.~Zhang, C.~S.~Li and Z.~Li,
  Phys.\ Rev.\  D {\bf 76} (2007) 116003
  [arXiv:0707.2132 [hep-ph]].

\bibitem{arXiv:0707.2451}
  Y.~Nakayama,
  Phys.\ Rev.\  D {\bf 76} (2007) 105009
  [arXiv:0707.2451 [hep-ph]].

\bibitem{arXiv:0707.2959}
  N.~G.~Deshpande, X.~G.~He and J.~Jiang,
  Phys.\ Lett.\  B {\bf 656} (2007) 91
  [arXiv:0707.2959 [hep-ph]].

\bibitem{hep-ex/0609040}
  A.~Abulencia {\it et al.}  [CDF Collaboration],
  Phys.\ Rev.\ Lett.\  {\bf 97} (2006) 242003
  [arXiv:hep-ex/0609040].

\bibitem{arXiv:0802.2255}
  V.~M.~Abazov {\it et al.}  [D0 Collaboration],
  arXiv:0802.2255 [hep-ex].

\bibitem{arXiv:0712.2397}
  T.~Aaltonen {\it et al.}  [CDF Collaboration],
  arXiv:0712.2397 [hep-ex].

\bibitem{arXiv:0803.0659}
  M.~Bona {\it et al.}  [UTfit Collaboration],
  arXiv:0803.0659 [hep-ph].

\bibitem{Lenz:2006hd}
  A.~Lenz and U.~Nierste,
  JHEP {\bf 0706} (2007) 072
  [arXiv:hep-ph/0612167].

\bibitem{arXiv:0707.0636}
  M.~Bona {\it et al.}  [UTfit Collaboration],
  JHEP {\bf 0803} (2008) 049
  [arXiv:0707.0636 [hep-ph]].

\bibitem{Becirevic:2001jj}
  D.~Becirevic {\it et al.},
  Nucl.\ Phys.\  B {\bf 634} (2002) 105
  [arXiv:hep-ph/0112303].

\bibitem{Deshpande:2008ra}
  N.~G.~Deshpande and X.~G.~He,
  arXiv:0806.2009 [hep-ph].

\bibitem{Fox:2007sy}
  P.~J.~Fox, A.~Rajaraman and Y.~Shirman,
  Phys.\ Rev.\  D {\bf 76} (2007) 075004
  [arXiv:0705.3092 [hep-ph]].

\bibitem{Buras:1990fn}
  A.~J.~Buras, M.~Jamin and P.~H.~Weisz,
  Nucl.\ Phys.\  B {\bf 347} (1990) 491.

\bibitem{Cabibbo:1963yz}
  N.~Cabibbo,
  Phys.\ Rev.\ Lett.\  {\bf 10}, 531 (1963).

\bibitem{Kobayashi:1973fv}
  M.~Kobayashi and T.~Maskawa,
  Prog.\ Theor.\ Phys.\  {\bf 49}, 652 (1973).

\bibitem{:2008vn}
   {\it et al.}  [Tevatron Electroweak Working Group and CDF Collaboration
                  and D0 Collab],
  arXiv:0808.1089 [hep-ex].

\bibitem{Inami:1980fz}
  T.~Inami and C.~S.~Lim,
  Prog.\ Theor.\ Phys.\  {\bf 65} (1981) 297
  [Erratum-ibid.\  {\bf 65} (1981) 1772].

\bibitem{Grinstein:2008qk}
  B.~Grinstein, K.~A.~Intriligator and I.~Z.~Rothstein,
  Phys.\ Lett.\  B {\bf 662}, 367 (2008)
  [arXiv:0801.1140 [hep-ph]].

\bibitem{Nakayama:2007qu}
  Y.~Nakayama,
  Phys.\ Rev.\  D {\bf 76} (2007) 105009
  [arXiv:0707.2451 [hep-ph]].

\bibitem{Mack:1975je}
  G.~Mack,
  Commun.\ Math.\ Phys.\  {\bf 55} (1977) 1.

\bibitem{next:unpart}
  J.~K.~Parry,
  arXiv:0810.0971 [hep-ph].

\end{thebibliography}
\end{document}